\documentstyle[12pt,aaspp4,graphicx]{article}
\newcommand{\bfig}{\begin{minipage}{3.3in}}
\newcommand{\efig}{\bigskip\end{minipage}}

\begin{document}

\title{Gamma-Ray Burst Afterglows in Pulsar-Wind Bubbles}


\author{Arieh K\"onigl$^1$ and Jonathan Granot$^2$}

\affil{$^1$Department of Astronomy \& Astrophysics and
Enrico Fermi Institute, University of Chicago, 5640 S. Ellis
Ave., Chicago, IL 60637; arieh@jets.uchicago.edu  \newline
$^2$Institute for Advanced Study, Olden Lane, Princeton, NJ
08540; granot@ias.edu}

\begin{abstract}
We propose to identify pulsar-wind bubbles (PWBs) as the
environment in which the afterglow emission in at least some
gamma-ray burst (GRB) sources originates. Such bubbles could
naturally account for both the high fraction of the internal
energy residing in relativistic electrons and positrons ($\epsilon_e$) and
the high magnetic-to-internal energy ratio ($\epsilon_B$) that
have been inferred in a number of sources from an interpretation of the
afterglow emission as synchrotron radiation. GRBs might occur
within PWBs under a number of scenarios: in particular,
in the supranova model of GRB formation a
prolonged (months to years) period of intense pulsar-type
wind from the GRB progenitor precedes the burst. Focusing
on this scenario, we construct a simple model of the
early-time structure of a plerionic supernova remnant (SNR), guided
by recent results on the Crab and Vela SNRs. 
The model is based on the assumption of an ``equipartition'' upper
bound on the electromagnetic-to-thermal pressure ratio in the
bubble and takes into account synchrotron-radiation cooling. We argue that the
effective upstream hydrogen number density for a relativistic
shock propagating into the bubble is
given by $n_{\rm H, equiv}= [4p +
(B^\prime+{\mathcal{E}}^\prime)^2/4\pi]/ m_p c^2$, where
$B^\prime$ and ${\mathcal{E}}^\prime$ are, respectively, the
comoving magnetic and electric fields and $p$ is the particle pressure.
We show that, for plausible parameter
values, $n_{\rm H, equiv}$ spans the range inferred from
spectral fits to GRB afterglows and that its radial profile
varies within the bubble and may resemble a uniform interstellar
medium or a stellar wind. We consider how
the standard expressions for the characteristic
synchrotron spectral quantities are modified when the
afterglow-emitting shock propagates inside a PWB instead of in a
uniform interstellar medium and demonstrate that the predictions
for the empirically inferred values of $\epsilon_e$ and
$\epsilon_B$ are compatible with the observations. Finally, we outline a
self-consistent interpretation of the X-ray emission features detected in
sources like GRB 991216 in the context of the supranova/PWB picture.
\end{abstract}



\section{Introduction}
\label{introduction}

Gamma-ray burst (GRB) sources are commonly interpreted in terms
of nonthermally emitting shocks associated with relativistic (and
possibly highly collimated) outflows from stellar-mass black
holes or strongly magnetized and rapidly rotating neutron stars
(see, e.g., Piran 1999 and M\'esz\'aros 2001 for
reviews). The prompt high-energy emission is thought to
originate in the outflow itself, with the $\gamma$-rays attributed to internal
shocks within the flow and with the associated optical ``flash'' and radio
``flare'' emission ascribed to the reverse shock that is driven
into the outflowing material as it starts to be decelerated by the inertia
of the swept-up ambient gas. By contrast, the longer-term, lower-energy
afterglow emission (see, e.g., \cite{VKW00} for a review) is
attributed to the forward shock that
propagates into the ambient medium. The ambient gas is usually
taken to be either the interstellar medium (ISM) of the host galaxy or
a stellar wind from the GRB progenitor star. 

It appears that most of the observed emission from GRBs and their
afterglows represents synchrotron radiation (e.g.,
\cite{SPN98}; \cite{PM98}; \cite{SP99}; \cite{CL00};
\cite{LP00}).
In view of source-energetics considerations, the
emission efficiency must be high. This implies that the
ratio $\epsilon_e$ of the internal energy in relativistic
electrons and positrons to the total internal energy density in the
emission region is not much smaller than 1, and that the ratio
$\epsilon_B$ of the magnetic-to-internal energy densities is not
much smaller than $\epsilon_e$. If the shocked gas consists of
protons and electrons, then only moderately high ($\lesssim
0.1$) values of $\epsilon_e$ may be expected even under
optimal circumstances (e.g., \cite{BM96}; \cite{PS00}). 
For $\epsilon_e$ to approach 1, it is probably necessary for the preshock gas 
to be composed primarily of $e^\pm$ pairs. A pair-dominated outflow is,
in fact, a feature of certain GRB models (e.g., \cite{U94};
\cite{MR97}; \cite{GW98}). Furthermore, the radiative efficiency of the
reverse shock (and possibly also of the forward shock during the early
afterglow phase) could be enhanced through pair creation
by the high-energy photons comprising the
gamma-ray pulse (e.g., \cite{TM00}; \cite{MRR01}). There is,
however, no natural way to account for large values of
$\epsilon_e$ during the later phases of afterglows in a typical
ISM or stellar-wind environment.

It is in principle also possible to account for comparatively
large values of $\epsilon_B$ in internal and
reverse shocks by appealing to shock compression of magnetized
outflows (e.g., \cite{SDD01}; \cite{GK01}, hereafter
GK). However, in the case of
afterglows in the standard scenario, the highest values of
$\epsilon_B$ that might be attained in this fashion (e.g., in a
shock propagating into a magnetized wind from a progenitor star;
see \cite{BC93}) could at best account only for the low end of the actual
range inferred in GRB afterglows ($\epsilon_B \gtrsim 10^{-5}$;
e.g., \cite{PK02}). For example, one could not explain in this
way the estimate $\epsilon_B \sim 0.01-0.1$ (derived by
model fitting of one of the most comprehensive spectral data sets
obtained to date) in the GRB 970508 afterglow (e.g., \cite{WG99};
\cite{GPS99}; \cite{CL00}).\footnote{The inferred value of $\epsilon_e$ in
this source is also fairly high ($\sim 0.1-0.6$).} 

As an alternative to compressional amplification of a preshock field,
various proposals have been advanced for generating strong
magnetic fields in the shocks themselves, but it is still unclear whether 
any of them can naturally account for a source like GRB 970508. 
For example, Medvedev \& Loeb (1999) suggested that a two-stream instability
behind the shock can generate fields that fluctuate on the very
short scale of the plasma skin depth. However, the most likely
value of $\epsilon_B$ predicted by this scheme is
rather low ($\ll 0.01$), as is also the value of $\epsilon_e$
(e.g., \cite{PS00}); furthermore, questions have been raised about
whether the fields will not, in fact, be damped on a similar microscopic
scale (Gruzinov 1999). Thompson \& Madau (2000) suggested that acceleration
of the preshock gas by the prompt gamma-ray pulse photons
would induce shearing motions that could significantly amplify the ambient
magnetic field. It is, however, unlikely that the preshock optical depth 
would be large enough for this effect to play a role for the
comparatively large spatial scales ($\gtrsim 10^{17}\ {\rm cm}$) and low
preshock densities ($\sim 0.03-3\ {\rm cm}^{-3}$) inferred for
the GRB 970508 afterglow (see M\'esz\'aros et
al. 2001). Blandford (2000), arguing by analogy with supernovae
like Cas A, proposed that the afterglow emission in a source
like GRB 970508 arises near the contact discontinuity that
separates the swept-up ambient gas from the outflowing matter, where these
two components can mix and interact. The large inferred magnetic
field presumably originates in the central source and undergoes
additional amplification in the turbulent interaction zone, but
a quantitative model of this scenario has not yet been
presented.\footnote{A related idea was discussed by Smolsky \&
Usov (2000), who considered a magnetized, pulsar-type wind and
suggested that it does not initially form a forward shock but
rather that the oscillating currents in the wind front excite large-amplitude
electromagnetic waves in the ambient medium and that high-energy
electrons accelerated in the front radiate in the field of these waves.
However, these authors still attributed afterglow emission
detected more than a day after the gamma-ray pulse (as was
the case for GRB 970508) to a conventional forward shock that develops ahead of
the wind front by that time. Additional ideas on how
large-amplitude electromagnetic waves in Poynting flux-dominated
outflows could lead to large values of $\epsilon_B$ in
afterglows were outlined by Lyutikov \& Blackman (2001).}

In this paper we propose that the large values of $\epsilon_B$
and $\epsilon_e$ inferred in afterglows like GRB 970508 arise
naturally if the outflow that gives rise to the gamma-ray pulse
expands into a pulsar-wind bubble (PWB). Such a bubble forms when the
relativistic wind (consisting of relativistic particles and
magnetic fields) that emanates from a pulsar shocks against the
ambient gas and creates a ``pulsar nebula,'' whose structure is
analogous to that of a stellar wind-inflated ``interstellar
bubble.'' When a bubble of this type expands inside a supernova
remnant (SNR), it gives rise to a ``plerionic'' SNR, of which
the Crab and Vela remnants are prime examples (see, e.g.,
Chevalier 1998 for a review). GRBs can arise inside PWBs
under a number of plausible scenarios, some of which have
already been considered in the literature. For example, Gallant
\& Achterberg (1999) suggested that, if GRB outflows are formed in
neutron-star binary mergers and expand into PWBs created by the
progenitor pulsars, then acceleration of relativistic ions in
the nebula by the forward shock could in
principle account for the observed population of
ultra-high-energy cosmic rays (UHECRs).\footnote{The association of
UHECRs with GRBs was originally proposed by
Waxman (1995), Milgrom \& Usov (1995), and Vietri (1995). Some
difficulties with
the simplest formulation of this idea were recently discussed by
Stecker (2000) and Farrar \& Piran (2000). UHECRs may, however,
originate in winds from young, rapidly spinning and strongly
magnetized neutron stars even if the latter are not linked to GRBs
(see \cite{BEO00}).} 
Vietri \& Stella (1998, 1999) presented a scenario for the
origin of GRBs in which a rotationally supported ``supramassive''
neutron star (SMNS) forms either by a supernova
explosion that is triggered by the collapse of a massive star or
as a result of mass
transfer in a low-mass X-ray binary (LMXB). In this picture, the
neutron star loses angular momentum (and associated centrifugal
support) through a pulsar-type wind until (on
a time scale of several months to several years) it becomes
unstable to collapse to a black hole
(a process that, in turn, induces a GRB outflow). Vietri \&
Stella (1998, hereafter VS) noted the analogy between the
proposed ``supranova''
remnants and plerionic SNRs, but they did not explicitly address
the structure of SMNS wind nebulae and their implications to GRB
afterglows.\footnote{In a recent paper, \cite{IGP01} discussed
some observational consequences of the precursor plerion in the
supranova scenario as well as the prospects for its direct
detection. Their picture differs from ours in that they assume
that the PWB disrupts and penetrates through the supernova
ejecta shell, attaining a size that is about an order of magnitude
larger than the SNR radius, 
and they associate the afterglow-emitting gas with entrained SNR
fragments or the ambient ISM. By contrast, in our picture the
PWB remains largely confined within the SNR shell, and the
afterglow emission arises in the high-effective-density interior
gas.} The afterglow sources observed to date are associated with
``long'' bursts (of duration $\gtrsim 2\ {\rm s}$) and are often
found within the projected image of the host galaxy. Such
sources could plausibly arise in the collapse of (or the merger
of a compact object with) a massive star (e.g., Woosley 2000),
although an LMXB progenitor may also be consistent with the data (\cite{VS99}).
In view of the evidence that
at least some afterglow sources are located along the line of
sight to a star-forming region (e.g., M\'esz\'aros 2001), we
adopt the supranova version of the SMNS scenario (VS)
as the underlying framework for our discussion.\footnote{It has
not yet been explicitly demonstrated that the supranova scenario can
account for long bursts; in fact, it has even been suggested
(\cite{BF01}) that this model is most likely to produce short
bursts. We note, however, that long bursts could in principle be
generated in the course of the collapse of the SMNS to a black
hole (see \cite{KR98}). Alternatively, if (as suggested by
VS) the GRB outflow is produced after the collapse in a
magnetized debris disk formed by the outer layers of the SMNS, then a long
duration could be a consequence of a comparatively low disk
viscosity (see, e.g., \cite{PWF99} and \cite{RJ99}) or of a
magnetically mediated spin-up torque exerted by the black hole
(\cite{VPO01}).} In \S 2 we estimate the physical
parameters of SMNS winds and of supranova remnants in light of
recent work on plerions, and we then model the structure of the
resulting PWBs. In \S 3 we consider the expected properties of
GRB afterglows that originate in such an environment. Our conclusions are
summarized in \S 4.

\section{Pulsar-Wind Bubbles in Young Supernova Remnants}
\subsection{The Supranova Scenario}
\label{scenario}

Supramassive neutron stars are general-relativistic equilibrium
configurations of rapidly rotating neutron stars whose masses
exceed the maximum mass of a nonrotating neutron star (e.g.,
\cite{CST94}; \cite{S94}). A uniformly rotating SMNS that loses
energy and angular momentum adiabatically while conserving its total baryon 
mass follows an evolutionary sequence that
brings it to a point where it becomes unstable to axisymmetric
perturbations, whereupon it undergoes a catastrophic collapse to
a black hole. In their supranova model, VS
postulated that the SMNS, which forms in the course of a
supernova explosion of a massive star, is magnetized and loses
energy and angular momentum through a pulsar-type
wind.\footnote{As noted by VS, gravitational radiation, possibly
associated with the excitation of $r$ modes in the SMNS (e.g., \cite{A98}),
is an alternative loss mechanism. However, in view of the
uncertainties involved in quantifying this process, we
follow VS and neglect it in the ensuing discussion.} The
rate of energy loss can be estimated from the magnetic dipole-radiation
formula
\begin{equation}\label{L_w}
L_w = {B_*^2R_*^6\Omega_*^4\over 6 c^3} = 7.0\times
10^{44} \left ({B_* \over 10^{12}\ G}\right )^2 \left({R_* \over
15\, {\rm km}}\right )^6 \left ({\Omega_* \over 10^4 \ {\rm
s}^{-1}} \right )^4\  {\rm ergs\ s^{-1}}\, ,
\end{equation}
where $B_*$ is the polar surface magnetic field, $R_*$ is the
circumferential radius (neglecting the distinction between its
equatorial and polar values in this approximation), and $\Omega_*$ is
the (uniform) angular velocity (whose maximum value is
$\sim 2\times 10^4\ {\rm s}^{-1}$; e.g., \cite{HLZ99}).\footnote{The 
magnetic dipole luminosity also scales as $\sin^2{\theta_{\Omega B}}$, 
where $\theta_{\Omega B}$ is the angle between the rotation and dipole
axes. However, as the spin-down torque of radio pulsars
is evidently largely independent of the value of $\theta_{\Omega B}$ 
(e.g., \cite{BS95}), we did not include this factor in equation
(\ref{L_w}).} The magnetic field amplitude in this estimate is 
normalized by
the typical radio-pulsar value. This situation is to be distinguished
from scenarios in which a magnetized rotator with a much
stronger field ($\gtrsim 10^{15}\ {\rm G}$) is
invoked to account for the GRB outflow itself (e.g., \cite{U94};
\cite{T94};  \cite{BY98}; \cite{KR98}; \cite{S99}; \cite
{RTK00}). The initial neutron-star magnetic field might be
amplified to such strengths by differential rotation (e.g.,
\cite{KR98}\footnote{These authors also mention the
possibility, suggested independently by VS
in their outline of the supranova scenario, that differential rotation
leading to a very strong field and possibly a GRB outflow could
be induced in an SMNS when it starts to collapse after losing
centrifugal support.}) or through dynamo action (e.g.,
\cite{TD93}\footnote{These authors associate the dynamo action
with neutrino flux-driven convection, which should occur if the
stellar spin period $2\pi/\Omega$ is shorter than the convective
overturn time $\sim 10^{-3}F_{39}^{-1/3}\ {\rm s}$ at the base
of the convection zone, where $F_{39}$ is the neutrino
heat flux in units of $10^{39}\ {\rm ergs\  cm^{-2}\ s^{-1}}$,
the expected value for a typical supernova explosion. Since
the energy released in a supranova explosion may be
significantly larger than in a typical supernova, and since the
neutron-star mass is also higher in
this case, the neutrino flux could be similarly larger
and the convection condition might not be satisfied, which would be
consistent with the assumption that the SMNS magnetic field does not
exceed typical pulsar values.}). For the comparatively low field amplitudes
adopted in the supranova scenario, the dynamical effect of the
magnetic field on the SMNS structure should be negligible (see,
e.g., \cite{B95}).

The wind power $L_w$ consists of electromagnetic and
particle contributions. The magnetic field is expected to have a
dominant toroidal component, which scales with distance $r$ from
the center as $1/r$. Under ideal-MHD conditions, the
Poynting-to-particle energy flux ratio in the wind is given by
\begin{equation}\label{sigma}
\sigma_w = {B_w^2 \over 4\pi\rho_w c^2}\ ,
\end{equation}
where $B_w$ is the field amplitude and $\rho_w$ is the rest-mass
density (both measured in the fluid frame), and it remains constant
along the flow after the terminal speed is reached.
There has been a great deal of debate
in the literature about the value of $\sigma_w$ in relativistic
pulsar outflows and about whether an ideal-MHD description is
appropriate (see, e.g., Arons 1998 and \cite{CLB98} for
discussions of this topic). For example, dynamical and radiative
models of the Crab pulsar
nebula have yielded preshock values of $\sigma_w$ in the range
$\sim 2-5 \times 10^{-3}$ (e.g., \cite{A02}).\footnote{Begelman
(1998), however, argued that a key underlying assumption of some
of the dynamical estimates --- that the magnetic field inside the shocked-wind
bubble maintains a coherent, large-scale, toroidal structure ---
may not be valid, and he suggested that $\sigma_w$ could be
higher in this source.} On the other hand, recent X-ray
observations of the Vela pulsar nebula have been interpreted
as implying $\sigma_w \sim 1$ (\cite{HGH01}).\footnote{This
interpretation was, however, questioned by Arons (2002), who suggested
that $\sigma_w$ in the Vela nebula is, instead, $<0.05$; an even lower upper
limit was given by Chevalier (2000), who proposed that
$\sigma_w< 10^{-4}$ in this source.} In view of
this uncertainty, and in order to examine the
dependence of our model results on this parameter, we derive
solutions for $\sigma_w$ in the range $10^{-3}-1$.

We also need to specify the Lorentz factor $\gamma_w$ and
the composition of the outflow. Spectral (e.g., \cite{KC84})
and optical brightness-distribution (\cite{GA94}) models of the Crab nebula
have implied a current value of $\sim 3 \times 10^6$ for
$\gamma_w$ upstream of the shock. However, although these fits account
for the optical--through--gamma-ray observations, they do not explain
the measured radio spectrum. In a recent model, Atoyan (1999)
interpreted the latter as being produced by a relic population
of relativistic electrons that had been accelerated during the early years
of the pulsar and that have subsequently lost most of their
energy by radiation and adiabatic-expansion losses. Based on
this interpretation, he argued that the Crab pulsar was born
with a period of 
$\sim 3-5\ {\rm ms}$ (as compared with previous
estimates of $\sim 19\ {\rm ms}$)\footnote{Independent arguments for why radio
pulsars like the Crab and Vela were likely born with rotation
periods $\lesssim 1\ {\rm ms}$ were recently given by Lai,
Chernoff, \& Cordes (2001) in the context of an interpretation of the
apparent alignment of the spin axes, proper motion directions,
and polarization vectors of the Crab and Vela pulsars.} and initially had 
$\gamma_w\le 10^4$. In light of this result, we adopt $\gamma_w = 10^4$ as a
fiducial value in our calculations: we assume that its magnitude is
roughly the same in all objects and that it does not change
significantly over the SMNS spin-down time. The pulsar outflow could consist of
$e^\pm$ pairs as well as ions. In fact, by modeling the wind
termination shock in the Crab nebula, Gallant \& Arons (1994)
inferred that the energy flux in ions is approximately twice
that in pairs in that source, and we already mentioned (see \S
1) the suggestion by Gallant \& Achterberg (1999) that UHECRs might be
identified with heavy ions in GRB-associated PWBs. Nevertheless, for
simplicity, we assume in our model that the SMNS wind is
composed purely of $e^\pm$ pairs.\footnote{If the ion and pair
energy fluxs are indeed comparable and these two components do
not exchange energy efficiently, then the resulting bubbles
would be approximated by the weakly cooling PWB solutions
presented in \S \ref{results}.} In this case the wind power can be written as
\begin{equation}\label{Lsigma}
L_w = 4\pi(1+\sigma_w)r^2n_w(r)\gamma_w^2\beta_wm_ec^3\, ,
\end{equation}
where $n_w(r)$ is the fluid-frame wind particle density at a
radius $r$, $m_e$ is the electron mass, and $\beta_wc$ is the
wind speed.\footnote{In view of
the large estimated value of $\gamma_w$, we henceforth set
$\beta_w$ equal to 1 in all numerical expressions.}

The spin-down time of a rapidly rotating SMNS can be estimated as
\begin{equation}\label{t_sd}
t_{\rm sd} = {\Delta E_{\rm rot} \over L_w} \approx 6 \left
( {\alpha \over 0.5}\right ) \left ({M_* \over 2.5 \, M_\odot}\right )^2
\left ( {R_* \over 15 \ {\rm km}} \right )^{-6} \left ({\Omega_*
\over 10^4\ {\rm s}^{-1}}\right )^{-3} \left ({B_* \over 10^{12}\
{\rm G}}\right )^{-2}\ {\rm yr}
\end{equation}
(see VS), where $\Delta E_{\rm rot} = \alpha G M_*^2 \Omega_*/2
c$ is the portion 
of the rotational energy of an SMNS of mass $M_*$ and angular velocity 
$\Omega_*$ that needs to be lost before it becomes unstable to
collapse.\footnote{The total rotational energy of the SMNS is given by 
$E_{\rm rot} = j G M_*^2 \Omega_*/2 c$, where the parameter $j$ measures
the stellar angular momentum in units of $GM_*^2/c$ and has values in the
range $0.57-0.78$ for realistic equations of state (e.g., Cook et
al. 1994; \cite{S94}).} The basic time scale is determined by the
underlying physical picture of a magnetized neutron star in
which a significant fraction of the binding energy is invested in
rotation (which is uniform, and thus does not lead to
field amplification much above typical pulsar values). However,
the expected variations in the parameter values that appear in
equation (\ref{t_sd}) could cause $t_{\rm sd}$ 
to range between a few months and a few years. It is instructive
to compare these values with Atoyan's (1999) estimate (obtained
from a fit to the Crab radio data) of the
initial spin-down time of the Crab pulsar, $t_{\rm sd}
\le 30\ {\rm yr}$ (a factor $\gtrsim 20$ smaller than previous
estimates that assumed a fixed functional dependence of the
spin-down torque on $\Omega$). The similarity of these estimates
is consistent with the possibility that the same modeling
framework may apply to both plerionic SNRs and SMNS-driven bubbles. 

Atoyan (1999) suggested
that the initial rotation energy of the Crab pulsar was
comparable to that of the supernova explosion that produced it,
and noted that his inferred value of $E_{\rm rot}$ ($\gtrsim 10^{51}\
{\rm ergs}$) was consistent with independent arguments (\cite{C77})
that the Crab nebula had originated in a normal Type II
supernova event. In the case of an SMNS it is, however, unlikely
that the explosion energy is as large as the
initial rotation energy ($\sim 10^{53}\ {\rm ergs}$), but since the
energy ($\Delta E_{\rm rot}$) deposited in the PWB is evidently
of the same order as $E_{\rm rot}$, the supernova ejecta
(subscript \textsf{ej}) would be accelerated by the expanding bubble and
one could obtain an approximate equality between $E_{\rm rot}$
and $E_{\rm ej} = 0.5\, M_{\rm ej} v_{\rm ej}^2$. For typical ejecta mass
$\gtrsim 10\ M_\odot$, this would imply $v_{\rm ej} \approx 0.1\, c$
at $t=t_{\rm sd}$ (about an order of magnitude higher than in a
typical SNR). This estimate of $v_{\rm ej}$ (which agrees with
that of VS) is supported by measurements of X-ray emission (e.g.,
\cite{P00}) and absorption (e.g., \cite{L01}) features in some
GRB sources (see \S \ref{features}).

In the supranova scenario, the GRB is associated with
the collapse of the SMNS, which occurs at a time $t_{\rm
sd}$ after the supernova explosion. Unless the explosion
takes place within a dense molecular cloud, the mass of the
swept-up ambient medium will remain negligible in comparison with
$M_{\rm ej}$ over this time scale and will not affect the
SNR dynamics. To simplify the treatment, we assume that this is
the case.\footnote{\label{evaporation}As a
further simplification, we neglect the possible incorporation of
mass from the ejecta shell into the bubble interior through evaporation
by the ``hot'' shocked-wind material. This effect, which has
been considered in the study of interstellar bubbles (e.g.,
\cite{W77}), would be strongly suppressed if the magnetic field
were strictly transverse to the flow direction, as is assumed
in our model. However, even
a small mean radial field component might lead to a thermal
conductivity that is high enough to significantly affect the
mass budget inside the bubble.} The expanding PWB
is expected to compress the ejecta into a thin shell and
accelerate it (e.g., \cite{RC84}). To within factors of order 1,
the outer radius of the bubble at time $t_{\rm sd}$ can be
approximated by the product of $v_b\equiv v_{\rm ej}(t_{\rm
sd})$ times the SMNS spin-down time:
\begin{equation}\label{R_b}
R_b = v_b t_{\rm sd} = 9.5 \times 10^{16} \beta_{b,-1}
\tau_{\rm sd}\ \ {\rm cm}\ ,
\end{equation}
where we set $v_b/c \equiv \beta_b = 0.1 \beta_{b,-1} $ and $t_{\rm sd}= \tau_{\rm sd}\ 
{\rm yr}$. To the extent that $v_b \propto (\Delta E_{\rm rot}/M_{\rm
ej})^{1/2}$ has nearly the same value in all sources, the
magnitude of $R_b$ is determined by that of $t_{\rm
sd}$. In a similar vein, if the energy $\Delta E_{\rm rot}=10^{53}\Delta E_{53}\
{\rm ergs}$ lost during the SMNS lifetime is approximately
constant from source to source ($\Delta
E_{53} \sim 1$), then $t_{\rm sd}$ can also be used to
parameterize the SMNS wind power: $L_w = \Delta E_{\rm rot}/t_{\rm
sd}= 3.2 \times 10^{45}\Delta E_{53}/\tau_{\rm sd}\ {\rm ergs\
s^{-1}}$. 

In their original proposal, VS focused on
the expected effect of the supranova ejecta and SMNS energy release
on the baryon content of the environment in which the GRB
occurs. This was motivated by the general requirement (see,
e.g., Piran 1999) that the burst energy be channeled
into a region with a relatively low number of massive particles
in order for the outflow to attain the high ($\gtrsim 10^2$)
Lorentz factors inferred in GRBs. However, this property of the
GRB outflow is probably determined primarily by the generic
properties of the central object (e.g., \cite{MR97};
\cite{KR98}; \cite{VK01})
rather than by the matter-sweeping action of the ejecta and SMNS
wind. Instead of this aspect of the supranova
scenario, we emphasize here the favorable consequences of the
expected delay between the
supranova explosion and the GRB event to the creation of
PWBs in which afterglows with high inferred values
of $\epsilon_B$ and $\epsilon_e$ could naturally
arise.\footnote{Another potential implication of
this delay, which we consider in \S \ref{features}, involves the 
interpretation of the X-ray spectral features detected in some GRB sources.}

\subsection{Wind-Bubble Structure}
\label{structure}

We follow previous treatments of PWB structure (\cite{RG74}; \cite{KC84};
\cite{EC87}) in our assumptions about the basic morphology of
the bubble: we take it to be spherical, with an outer radius
$R_b$, and assume that the pulsar wind propagates freely (with
$\sigma_w={\rm const}$)
until it is shocked at a radius $R_s$. Our model differs, however,
from previous treatments in that we take account of
nonthermal radiation losses (which could be important during the early
phase of the nebula) and we do not assume that ideal MHD
is necessarily applicable throughout the shocked-wind bubble. As has been
demonstrated in the previously cited papers, a PWB that expands
adiabatically with a nonrelativistic speed and that contains a large-scale
toroidal magnetic field frozen into the matter
corresponds to $\sigma_w \approx \beta_b \ll 1$. Such a model
thus cannot describe a bubble with $\beta_b \ll 1$ and
$\sigma_w \lesssim 1$.\footnote{The presence of a large-scale
toroidal magnetic field also implies that the bubble will become
elongated and will assume a cylindrical, rather than a
spherical, shape (see \cite{BL92}). We return to the question of the bubble
morphology in \S \ref{features}.} If $\sigma_w\sim 1$, then the
postshock flow is magnetically dominated from the
start. But even if $\sigma_w
\ll 1$ and the postshock value of the fluid-frame magnetic-to-particle
pressure ratio $p_B/p$ is $< 1$, this ratio will grow with
radius $r$ in the bubble and, if radiative cooling is even
moderately important and flux freezing continues to hold, its value will
at some point increase above 1 and could eventually become $\gg
1$. However, as was already argued by Rees \&
Gunn (1974), a situation in which $p_B$ significantly exceeds
$p$ is unlikely to persist in a real PWB. We therefore adopt an
alternative formulation and drop the assumption of ideal MHD in
the shocked gas at the point where the electromagnetic pressure
first rises above the particle pressure. We assume, instead, that
beyond that point the electromagnetic pressure in the bubble remains in
approximate equipartition with the particle
pressure.\footnote{An equipartition assumption (between the
thermal and magnetic pressure components) was previously
incorporated as a limiting case in the (Newtonian)
plerion evolutionary model of Reynolds \& Chevalier (1984). 
Possible physical mechanisms for the
breakdown of ideal MHD when $p_B$ increases to $p$ were discussed by 
Kennel, Gedalin, \& Lominadze (1988), Begelman (1998), and
Salvati et al. (1998).}
For definiteness, we assume that the flow obeys ideal MHD within
the wind shock and we fix the electromagnetic-to-particle
pressure ratio $\delta\equiv ({\mathcal{E^\prime}}^2+{B^\prime}^2)/{8\pi p}$ in
the bubble by setting
\begin{equation}\label{delta}
\delta = {\rm
const} = \mathsf{max}\{\delta_{\rm ps}\, , \, 1\}\, , \quad\quad\quad
\mathsf{max}\{R_s\, ,\, R_{\rm eq} \}\le r \le R_b\, .
\end{equation}
Here $\mathcal{E^\prime}$ and $B^\prime$ are, respectively, the
fluid-frame electric and magnetic fields, the subscript \textsf{ps} denotes
postshock quantities, and $R_{\rm eq}$ is the radius where $p_B/p$ first 
increases to 1 (assuming ideal-MHD evolution) if $\delta_{\rm ps}<1$. 
According to this prescription, when the postshock value of $\delta$ 
is less than 1 ($\delta_{\rm ps}<1$), the bubble flow
starts out maintaining flux freezing, but if $p_B/p$ increases
to 1 before the outer boundary is reached, it switches to a
nonideal evolution (with $\delta$ fixed at $1$) beyond
$R_{\rm eq}$. If, however, $\sigma_w$ is large enough that
$\delta_{\rm ps}\geq 1$, then the entire
PWB volume is subject to the euipartition constraint, with
$\delta$ fixed at $\delta_{\rm ps}$.
For a strong shock, the postshock value of $\delta$ can be
expressed as a function of the wind magnetization parameter
$\sigma_w$ and speed $\beta_w$.\footnote{In typical applications, the
speed of the wind shock
is much lower than that of the wind, and even than $v_b$ (see \cite{EC87}). 
For the sake of simplicity, we therefore set it equal to zero and identify the
rest frame of the wind shock with that of the source.} Specifically,
\begin{equation}\label{delta_ps2}
\delta_{\rm ps}={{B_{\rm ps}^\prime}^2\over 8\pi p_{\rm
ps}}={4\gamma_{\rm ps}^2\beta_{\rm ps}^2+1\over
2\sigma_w^{-1}\gamma_{\rm ps}^2\beta_{\rm ps}^2-1+
(\gamma_{\rm ps}\beta_{\rm ps}/\gamma_w\beta_w)^2}\ ,
\end{equation}
where $\gamma_{\rm ps}\equiv (1-\beta^2_{\rm ps})^{-1/2}$ is the 
Lorentz factor of the postshock flow and $\beta_{\rm ps}$
is given by the solution of the equation
\begin{equation}\label{eq_beta_ps}
\beta_{\rm ps}^3-{\beta_w(4+5\sigma_w+2\sigma_w\gamma_w^{-2}\beta_w^{-2})\over 
3(\sigma_w+1)}\beta_{\rm ps}^2
+{\beta_{\rm ps}\over 3}+{\beta_w\sigma_w\over 3(\sigma_w+1)}=0\ .
\end{equation}
In the limit $\gamma_w \gg 1$, both $\delta_{\rm ps}$ and
$\beta_{\rm ps}$ become functions of $\sigma_w$ alone (see GK):
equation (\ref{delta_ps2}) simplifies to
\begin{equation}\label{delta_ps}
\delta_{\rm ps}={{B_{\rm ps}^\prime}^2\over 8\pi p_{\rm
ps}}={4\gamma_{\rm ps}^2\beta_{\rm ps}^2+1\over
2\sigma_w^{-1}\gamma_{\rm ps}^2\beta_{\rm ps}^2-1}\ ,
\end{equation}
whereas the equation for $\beta_{\rm ps}$ is reduced from a
cubic to a quadratic, with the solution
\begin{equation}\label{beta_ps}
\beta_{\rm
ps}={1+2\sigma_w+[16\sigma_w(1+\sigma_w)+1]^{1/2}\over 6(1+\sigma_w)}\ .
\end{equation}
For $\sigma_w = 10^{-3},\, 10^{-2},\, 0.1$, and 1, these expressions
yield $\{\beta_{\rm ps},\, \delta_{\rm ps}\} =
\{0.33,\,0.006\},\ \{0.35,\,0.059\},\ \{0.43,\,0.53\}$ and
$\{0.73,\,4.37\}$, respectively. Note that, for $\sigma_w \le 0.2$
(corresponding to $\beta_{\rm ps} \le 0.5$), $\delta_{\rm ps}$
is $\le 1$, so that (by eq. [\ref{delta}]) $\delta=1$.

We simplify the treatment of the bubble interior by assuming
that the flow is purely radial, that [in spherical coordinates
($r,\,\theta,\,\phi$)] the magnetic field continues to possess only a $\phi$
component (${\mathbf B} = B\, {\mathbf \hat \phi}$; see \cite{B98}) and the
electric field only a $\theta$ component (${\mathbf
{{\mathcal{E}}}}={{\mathcal{E}}}\, {\mathbf \hat \theta}$; this follows from
the previous two assumptions when ideal MHD is applicable, but
needs to be postulated when it is not), and that the only nonzero
spatial derivatives are in the radial direction. Under these
assumptions, the particle number, energy,
and momentum conservation equations in the PWB take the form
\begin{eqnarray}
&{1\over c}{\partial\over\partial t}(\gamma n)+{1\over
r^2}{\partial\over\partial r}(r^2\gamma \beta n)=0\, , \label{particle}
\\
&{1\over c}{\partial\over\partial t}\left(\gamma^2
w-p+{{\mathcal{E}}^2+B^2\over 8\pi}\right)+
{1\over r^2}{\partial\over\partial
r}\left[r^2\left(\gamma^2 \beta w+{{{\mathcal{E}}}B\over
4\pi}\right )\right]=-{\gamma\Lambda\over c}\ , \label{energy}
\\
&{1\over c}{\partial\over\partial t}\left(\gamma^2 \beta w+
{{\mathcal{E}}B\over 4\pi}\right )+{1\over r^2}{\partial\over\partial
r}\left[r^2\left(\gamma^2\beta^2w+{{{\mathcal{E}}}^2 + B^2\over 8\pi}
\right)\right]+{\partial p\over\partial r}= -{\gamma\beta\Lambda\over
c}\label{momentum}\  ,
\end{eqnarray}
where $n$ is the particle number density, $w=\rho c^2+e+p$ is the
enthalpy density (with $e$ being the internal energy density and
$\rho$ the rest-mass density), $\Lambda$ is the emissivity
(which, like the preceding quantities, is measured in the
fluid rest frame), $\beta$ is the radial speed in units of $c$,
and $\gamma$ is the Lorentz factor. The electric and magnetic
field amplitudes in these equations are measured in the
central-source frame; they are related to their fluid-frame
counterparts through the Lorentz transformations
\begin{equation}\label{EB}
{{\mathcal{E}}}^\prime = \gamma({\mathcal{E}}-\beta B)\ , \quad\quad
B^\prime=\gamma(B-\beta {\mathcal{E}})\ .
\end{equation}
The evolution of ${{\mathcal{E}}}$ is governed by Faraday's law,
\begin{equation}\label{Faraday}
{1\over c} {\partial B\over \partial t} + {1\over r}{\partial
\over \partial r}\left ( r {{\mathcal{E}}} \right ) = 0\ .
\end{equation}

Given that $\gamma_w \gg 1$, the shocked
gas should be well described by a relativistic equation of state
\begin{equation}\label{rel_eos}
p={e\over 3}={w\over 4}\ .
\end{equation}
The acoustic speed $v_{\rm ac}$ would be correspondingly high. For
example, in the ideal-MHD regime, where it is given by
$[(1/3+\delta/2)/(1+\delta/2)]^{1/2}c$ (representing the phase speed
of the fast-magnetosonic wave; e.g., \cite{K80}), $v_{\rm ac}/c\approx
0.75$ for $\delta \approx 1$, which we take to be large enough
in comparison with $\beta_b$ 
($\approx 0.1$) to justify a steady-state approximation within the
bubble. We therefore set $\partial/\partial t=0$ in equations
(\ref{particle})--(\ref{Faraday}). Equation (\ref{particle})
then yields
\begin{equation}\label{particle2}
r^2\gamma\beta n=R_s^2\gamma_{\rm ps}\beta_{\rm ps}n_{\rm ps}\equiv C\, .
\end{equation}
The constant $C$ can be evaluated from the shock jump condition
\begin{equation}\label{particle3}
\gamma_{\rm ps}\beta_{\rm ps}n_{\rm ps} = \gamma_w\beta_wn_w\, ,
\end{equation}
where we have assumed that there is no pair production at the wind shock. 
Using also equation (\ref{Lsigma}), one obtains
\begin{equation}\label{C}
C= {L_w \over 4\pi(1+\sigma_w)\gamma_wm_ec^3}\ .
\end{equation}
Under the steady-state assumption, equation (\ref{Faraday})
implies that ${{\mathcal{E}}}(r)
\propto 1/r$ inside the PWB. Normalizing to the value
immediately behind the shock, we infer
\begin{equation}\label{Er}
{{\mathcal{E}}}(r) = \frac{{{\mathcal{E}}}_{\rm ps}R_{\rm ps}}{r}= \left (
\frac{\sigma_w}{1+\sigma_w} \frac{\beta_wL_w}{c} \right )^{1/2}
\frac{1}{r}\ .
\end{equation}

Elimination of the radiative cooling term from equations
(\ref{energy}) and (\ref{momentum}) leads to
\begin{equation}\label{dr1}
{dp\over dr}+\gamma^2\beta w{d\beta\over dr}+{B -
\beta{{\mathcal{E}}} \over 4\pi r}{d\over dr}
(r B)=0\, ,
\end{equation}
whereas subtraction of $\gamma\beta$ times equation (\ref{momentum}) from
$\gamma$ times equation (\ref{energy}) yields
\begin{equation}\label{entropy}
{1\over r^2}{d\over dr}(r^2 \gamma\beta w) - \gamma\beta{dp\over
dr} + {\gamma({\mathcal{E}}-\beta B)\over
4\pi r}{d\over dr}(rB)=-{\Lambda\over c}
\end{equation}
(the entropy equation), where in both cases we took account of
the constancy of the product $r{{\mathcal{E}}}$ inside the bubble.
If $\sigma_w$ is not $\ll 1$, then most of the bubble volume will
be in the equipartition regime ($\delta \approx 1$, or,
equivalently, $\epsilon_B\approx \epsilon_e$), in which case
$\Lambda$ will typically be dominated by synchrotron
radiation.\footnote{In the equipartition region, synchrotron
self-Compton emission cannot exceed the synchrotron
radiation under any circumstances: it is comparable to the
synchrotron emission if the bubble is highly radiative, but it
remains much smaller if the radiative cooling time is longer than the
bubble expansion time.} To simplify the treatment, we
take synchrotron emission to be the main radiative cooling
process even for low values of $\sigma_w$. Furthermore, we assume that at any
given location within the bubble the $e^\pm$ pairs have a monoenergetic energy
distribution characterized by a random (or ``thermal'') Lorentz factor
$\gamma_e$. The latter approximation is appropriate if the
postshock gas undergoes significant 
radiative cooling (e.g., \cite{GPS00}), which, as we discuss in
\S \ref{results}, may be the case in SMNS-driven bubbles. The
synchrotron emissivity can then be written in the form
\begin{equation}\label{Lambda}
\Lambda={4\over 3}\sigma_T c n \gamma_e^2 {{B^\prime}^2\over 8\pi}\ ,
\end{equation}
where $\sigma_T$ is the Thomson cross section. In view of
equation (\ref{rel_eos}), it is then also possible to write
the particle pressure as  
\begin{equation}\label{p_e}
p={1 \over 3} \gamma_e n m_e c^2\, .  
\end{equation}
Combining this expression with equation (\ref{particle2}) gives
\begin{equation}\label{gamma_e}
\gamma_e= Dr^2\gamma\beta p\, ,
\end{equation}
where $D \equiv 3/m_ec^2C=3/\gamma_{\rm ps}\beta_{\rm ps}n_{\rm
ps}R_s^2m_ec^2$.
Using equations (\ref{EB}), (\ref{C}), (\ref{p_e}), and (\ref{gamma_e})
in equation (\ref{Lambda}), one can
express the radiative cooling term in equation (\ref{entropy})
in the form
\begin{equation}\label{Lambda2}
{\Lambda\over c}= G \gamma^3\beta(rB-\beta r{\mathcal{E}})^2 p^2\, ,
\end{equation}
where 
$G\equiv (\sigma_T/2\pi m_ec^2)D=6\sigma_T(1+\sigma_w)\gamma_w/m_ecL_w$.

We now consider the term $rB$ that appears in equations (\ref{dr1}),
(\ref{entropy}), and (\ref{Lambda2}). Its form depends on
whether the flow is in the ideal-MHD or the equipartition regime
of the bubble interior. The ideal-MHD case corresponds to
setting ${\mathcal{E}}^\prime = 0$ in equation (\ref{EB}), which
implies $rB = r{\mathcal{E}}/\beta\propto 1/\beta$ (by
eq. [\ref{Er}]). It is then straigthtforward to obtain from
equations (\ref{dr1}) and (\ref{entropy}) (after also
substituting $w=4p$ from eq. [\ref{rel_eos}]) the following pair
of coupled, first-order, ordinary differential equations for the
variables $\beta$ and $p$, which give the structure of the
ideal-MHD sector of the PWB:
\begin{equation}
{d\beta\over dr}=
\left[Gr^2{\mathcal{E}}^2p^2+8\gamma^2\beta^2\frac{p}{r}\right
]\left[4\gamma^4\beta(3\beta^2-1)p
-\frac{3}{4\pi}\frac{{\mathcal{E}}^2}{\beta}\right]^{-1}\ , \ \
R_s\le r \le \mathsf{min}\{R_{\rm eq}\, ,\, R_b\}\,
,\label{beta_ideal}
\end{equation}
\begin{eqnarray}&&
{dp\over dr} = \left [ 4 \gamma^2\beta^2 p - \frac{{\mathcal{E}}^2}{4\pi
\gamma^2\beta^2}\right ] \left[Gr^2{\mathcal{E}}^2p^2 +
8\gamma^2\beta^2\frac{p}{r} \right ] \left [
\frac{3}{4\pi}{\mathcal{E}}^2 +
4\gamma^4\beta^2(1-3\beta^2)p\right ]^{-1} \nonumber \\ &&
\quad\quad\quad\quad\quad\quad\quad\quad\quad\quad
R_s\le r \le \mathsf{min}\{R_{\rm eq}\, ,\,
R_b\}\, , \label{p_ideal}
\end{eqnarray}
with ${\mathcal{E}}$ given by equation (\ref{Er}).

For the nonideal (equipartition) regime, we combine equations
(\ref{delta}) and (\ref{EB}) to obtain a quadratic equation for
$rB$, whose relevant root is
\begin{equation}\label{B_eq}
rB_{\rm eq} = \frac{2\beta}{1+\beta^2}r{\mathcal{E}}
+ \frac{[8\pi \delta \gamma^2 (1+\beta^2)r^2
p-r^2{\mathcal{E}}^2]^{1/2}}{\gamma^2(1+\beta^2)}\ .
\end{equation}
Using this relation as well as equations (\ref{rel_eos}),
(\ref{Er}), and (\ref{Lambda2}) in equations (\ref{dr1})
and (\ref{entropy}), it is once again possible to extract
explicit differential equations for $\beta$ and $p$:
\begin{eqnarray}
&&
{d\beta \over dr}=\left \{ \left [1+{\delta (rB_{\rm eq}-\beta
r{\mathcal{E}})\over \gamma^2[(1+\beta^2)rB_{\rm eq}-2\beta 
r{\mathcal{E}}]}\right ]G\gamma^2\beta (rB_{\rm eq}-\beta
r{\mathcal{E}})^2p^2 + \nonumber \right. \\ &&
\left. +\left [ 8\beta +\frac{2\delta r
{\mathcal{E}}}{\gamma^4[(1+\beta^2)rB_{\rm eq} -2\beta r
{\mathcal{E}}]}\right ]\frac{p}{r}\right \} \left \{
4 \left [\gamma^2
(3\beta^2-1)-\delta \right ] p + \right. \nonumber \\ &&
\left. +{(r^2{\mathcal{E}}B_{\rm eq}-8\pi\delta \gamma^2\beta r^2 p)
\left[4\beta rB_{\rm eq}-(1+3\beta^2)r{\mathcal{E}}\right] \over
2\pi(1+\beta^2)\gamma^2[(1+\beta^2)rB_{\rm eq}-2\beta 
r{\mathcal{E}}]r^2} \right \}^{-1}\, , \nonumber \\ &&
\quad\quad\quad\quad\quad\quad\quad\quad\quad\quad
\mathsf{max}\{R_s\, ,\, R_{\rm eq}\}\le r \le R_b\, ,\label{beta_eq}
\end{eqnarray}
\begin{eqnarray}
&&{dp\over dr} = \left \{1+\frac{\delta(rB_{\rm eq}-\beta
r{\mathcal{E}})}{\gamma^2[(1+\beta^2)rB_{\rm eq}-2\beta
r{\mathcal{E}}]}\right \}^{-1}\times \nonumber \\ &&
\times \left\{\left[{(\beta r {\mathcal{E}}-rB_{\rm
eq})(r^2{\mathcal{E}}B_{\rm eq}-8\pi\delta\gamma^2\beta r^2p)\over
2\pi(1+\beta^2)\gamma^2[(1+\beta^2)rB_{\rm eq}-2\beta r
{\mathcal{E}}]r^2}-4\gamma^2\beta p\right ] \left({d\beta \over dr}
\right ) + {2\delta(\beta r{\mathcal{E}}-rB_{\rm eq})\over 
\gamma^2[(1+\beta^2)rB_{\rm eq}-2\beta r{\mathcal{E}}]}
\frac{p}{r} \right \}\ , \nonumber \\ &&
\quad\quad\quad\quad\quad\quad
\mathsf{max}\{R_s\, ,\, R_{\rm eq}\}\le r \le R_b\, ,\label{p_eq}
\end{eqnarray}
where the term $(d\beta/dr)$ in equation (\ref{p_eq}) is given by
the expression (\ref{beta_eq}).
Equations (\ref{beta_ideal})--(\ref{p_eq}) are integrated over
their respective validity domains subject to
the boundary conditions
\begin{equation}\label{BC}
\beta(R_s)=\beta_{\rm ps}\, , \quad\quad p(R_s)=p_{\rm ps}\, ,
\end{equation}
where $\beta_{\rm ps}$ is given by equation (\ref{beta_ps}) and
the postshock pressure
\begin{equation}\label{p_ps}
p_{\rm ps} = {\sigma_w \gamma_w^2\beta_w^2n_w(R_s)m_ec^2
\over 2\delta_{\rm ps} \gamma_{\rm ps}^2\beta_{\rm ps}^2}
\end{equation}
(with $n_w(r)$ given by eq. [\ref{Lsigma}]) is similarly obtained from the
wind-shock jump conditions 
(see GK). The value of $R_s$, where the boundary conditions
(\ref{BC}) are imposed, is not known a priori and must be determined from an
additional constraint. This can be provided by requiring
global particle conservation: for a bubble considered at time
$t$ after the supranova explosion, the total number of particles
within the radius $R_b(t)$ [which consists of the unshocked wind
at $r<R_s(t)$ and the shocked wind at $r>R_s(t)$] is equal to the
total number of particles injected by the central neutron star over
the time $t$. The pair injection rate at the source is given by
\begin{equation}\label{Ndot}
\dot{N}= {L_w \over (1+\sigma_w)\gamma_w m_e c^2}\, ,
\end{equation}
and hence the total number of particles within $R_b$ at time $t$
is $N(t)=\dot{N}t$. 
We approximate $t\approx R_b/\beta_bc$, which should be accurate
to within a factor of order 1 (for example, $t=1.5R_b/\beta_bc$ in
the case of an adiabatic bubble, with the numerical coefficient decreasing
in the presence of cooling; see \cite{RC84}). The number of
particles within the volume
occupied by the unshocked wind is thus
\begin{equation}\label{N_1}
N(r<R_s)=\dot{N}{R_s\over\beta_w c}\approx N(t){R_s\over
ct}\approx \beta_bN{R_s\over R_b}\ ,
\end{equation}
whereas the total number of particles within the shocked-wind
region is
\begin{equation}\label{N_2}
N(R_s<r<R_b)=\int_{R_s}^{R_b} 4\pi r^2\gamma n dr=
{\beta_b N \over R_b}\int_{R_s}^{R_b} {dr\over\beta}\ ,
\end{equation}
where we used equations (\ref{R_b}), (\ref{particle2}),
(\ref{C}), and (\ref{Ndot}). The solutions obtained in this manner will not, in
general, be entirely self-consistent, since the bubble structure evolves with
time whereas we have assumed a steady state. The wind gas cannot
arrive at $r=R_b$ at the speed $v_b$ after traveling from the
origin for the same
duration (viz., the age of the bubble) as the ejecta that is currently at 
$R_b$, given that the wind speed is $>v_b$ everywhere within
this region and that the ejecta speed was $<v_b$ before it reached $R_b$.
This argument implies that, if the particle-conservation
condition is imposed, then $\beta(R_b)$ will be lower than (rather
than exactly match) $\beta_b$. 
This is not a serious inconsistency, since the flow near $R_b$
will generally be highly subsonic 
and therefore can readily adjust to match the speed ($v_b$) of
the outer boundary. Nevertheless, in order to assess the sensitivity of
the results to the choice of the imposed constraint, we also solve
the system of equations subject to the alternative condition
$\beta(R_b) = \beta_b$. 

\subsection{Illustrative Solutions}
\label{results}

The governing equations can be rendered
dimensionless by introducing $\tilde{r}\equiv r/R_b$ (with 
$\tilde{r}_s$ and $\tilde{r}_{\rm eq}$ denoting the dimensionless
counterparts of $R_s$ and $R_{\rm eq}$, respectively) as well as
$\tilde{B}\equiv B/{\mathcal{E}}(R_b)$ and
$\tilde{p}\equiv p/p_{\rm ref}$, where 
\begin{equation}\label{p_ref}
p_{\rm ref}\equiv
\frac{{{\mathcal{E}}}^2(R_b)}{8\pi}=\left
(\frac{2\sigma_w}{1+\sigma_w}\right ) \left (\frac{\beta_wL_w}
{16\pi R_b^2c}\right ) \equiv \left ( \frac{2\sigma_w}{1+\sigma_w}\right ) p_1\ ,
\end{equation}
(see eq. [\ref{Er}]).\footnote{The parameter $p_1$ is defined
so as to isolate the $\sigma_w$-independent part of $p_{\rm ref}$: it is equal to
$p_{\rm ref}$ evaluated at $\sigma_w=1$.} They then take the form 
\begin{equation}
{d\beta\over d\tilde{r}}
=\left[\frac{3a}{8\pi}\tilde{p}^2+4\gamma^2\beta^2\frac{\tilde{p}}
{\tilde{r}}\right ]\left [2\gamma^4\beta
(3\beta^2-1)\tilde{p}-\frac{3}{\beta\tilde{r}^2}\right ]^{-1}\ ,
\quad\quad\quad \tilde{r}_s\le \tilde{r} \le
\mathsf{min}\{\tilde{r}_\delta\, ,\, 1\}\, ,\label{beta_ideal1}
\end{equation}
\begin{eqnarray} &&
{d\tilde{p}\over d\tilde{r}} = \left [ 4 \gamma^2\beta^2
\tilde{p} - \frac{2}{\gamma^2\beta^2\tilde{r}^2}\right ]
\left[\frac{3a}{8\pi}\tilde{p}^2 +
4\gamma^2\beta^2\frac{\tilde{p}}{r} \right ]\times \nonumber \\ &&
\times \left[ \frac{3}{\tilde{r}^2} + 2\gamma^4\beta^2
(1-3\beta^2)\tilde{p}\right ]^{-1}\ ,
\quad\quad\quad \tilde{r}_s\le \tilde{r} \le
\mathsf{min}\{\tilde{r}_\delta\, ,\, 1\}\, ,\label{p_ideal1}
\end{eqnarray}
\begin{equation}
\tilde{r}\tilde{B}_{\rm eq} =
\frac{2\gamma^2\beta+[\delta\gamma^2(1+\beta^2)\tilde{r}^2\tilde{p}-1]^{1/2}}
{\gamma^2(1+\beta^2)}\ , \label{B_eq1}
\end{equation}
\begin{eqnarray} 
&&
{d\beta \over d\tilde{r}}=
\left \{ \left [1+\frac{\delta(\tilde{r}\tilde{B}_{\rm eq}-\beta)}
{\gamma^2[(1+\beta^2)\tilde{r}\tilde{B}_{\rm eq}-2\beta]}\right ]
\frac{3a}{16\pi}\gamma^2\beta(\tilde{r}\tilde{B}_{\rm
eq}-\beta)^2\tilde{p}^2 + \nonumber \right. \\ &&
\left. + \left [ 2\beta
+\frac{\delta}{2\gamma^4[(1+\beta^2)\tilde{r}\tilde{B}_{\rm eq}
-2\beta]}\right ]\frac{\tilde{p}}{\tilde{r}}\right \}
\left \{ \left [\gamma^2(3\beta^2-1)
-\delta \right ]\tilde{p} + \right. \nonumber \\ && 
\left. +{(\tilde{r}\tilde{B}_{\rm
eq}-\delta\gamma^2\beta\tilde{r}^2\tilde{p})(4\beta\tilde{r}\tilde{B}_{\rm
eq}-1-3\beta^2)
\over (1+\beta^2)\gamma^2[(1+\beta^2)\tilde{r}\tilde{B}_{\rm
eq}-2\beta ]\tilde{r}^2 } \right \}^{-1}\, , \nonumber \\ &&
\quad\quad\quad\quad\quad\quad
\mathsf{max}\{\tilde{r}_s\, ,\, \tilde{r}_\delta\}\le \tilde{r} \le 1\,
,\label{beta_eq1}
\end{eqnarray}
\begin{eqnarray}&&
{d\tilde{p}\over d\tilde{r}} = \left
\{1+\frac{\delta(\tilde{r}\tilde{B}_{\rm
eq}-\beta)}{\gamma^2[(1+\beta^2)\tilde{r}\tilde{B}_{\rm
eq}-2\beta]}\right \}^{-1}\times \nonumber \\ &&
\times \left\{4\left[{(\beta-\tilde{r}\tilde{B}_{\rm eq})
(\tilde{r}\tilde{B}_{\rm eq}-\delta\gamma^2\beta\tilde{r}^2\tilde{p})\over
(1+\beta^2)\gamma^2[(1+\beta^2)\tilde{r}\tilde{B}_{\rm eq}-2\beta]
\tilde{r}^2}-\gamma^2\beta \tilde{p}\right ] \left({d\beta \over d\tilde{r}}
\right ) + {2\delta(\beta-\tilde{r}\tilde{B}_{\rm eq})\over 
\gamma^2[(1+\beta^2)\tilde{r}\tilde{B}_{\rm eq}-2\beta]}
\frac{\tilde{p}}{\tilde{r}} \right \}\ , \nonumber \\ &&
\quad\quad\quad\quad\quad\quad
\mathsf{max}\{\tilde{r}_s\, ,\, \tilde{r}_\delta\}\le \tilde{r} \le 1\,
,\label{p_eq1}
\end{eqnarray}
(corresponding to eqs. [\ref{beta_ideal}]--[\ref{p_eq}],
respectively), where
\begin{equation}\label{a}
a\equiv \left [{\sigma_T L_w \over (1+\sigma_w)m_ec^3R_b}\right
] (\sigma_w\beta_w)^2\gamma_w\ \equiv \left (
\frac{2\sigma_w^2}{1+\sigma_w} \right ) a_1 \ .
\end{equation}
The parameter $a_1$ in equation (\ref{a}) was introduced so as
to isolate the $\sigma_w$-independent part of the parameter $a$:
it is equal to $a$ evaluated at $\sigma_w=1$.\footnote{Note that
the term in square parentheses in the expression for $a$ can
be interpreted as a ``bubble compactness parameter,'' using the
terminology often employed in studies of compact astrophysical
pair configurations.} 
As we explicitly demonstrate below, $a_1$ measures the
relative importance of radiative cooling within the bubble.
Numerically, $1/a_1$ is of the order of the nominal radiative cooling time
of the bubble in units of $R_b/c$, and hence the larger the
value of $a_1$, the stronger the role that radiative cooling plays in 
determining the bubble structure. In the supranova model, if $\beta_b$ 
and  $\Delta E_{\rm rot}$ are approximately constant from source to source,
then $a_1$ scales with the bubble age $t_{\rm sd}$ at
the time of the GRB as roughly $t_{\rm sd}^{-2}$ (see
eqs. [\ref{t_sd}] and [\ref{R_b}]).

The boundary conditions given by equation (\ref{BC}) are applied
at $\tilde{r}_s$, which, in turn, is given either by the relation 
\begin{equation}\label{BC1}
{\beta_b\over 1-\beta_b\tilde{r}_s}
\int_{\tilde{r}_s}^1
{d\tilde{r}\over\beta(\tilde{r})}= 1
\end{equation}
(see
eqs. [\ref{Ndot}]--[\ref{N_2}]) or by imposing the condition
\begin{equation}\label{BC2}
\beta(\tilde{r}=1)=\beta_b
\end{equation}
We solve this system of equations
for given choices of the parameter $\sigma_w$ and $a_1$
by iterating on the value of $\tilde{r}_s$ until both the boundary conditions 
(\ref{BC}) and the constraint (\ref{BC1}) or (\ref{BC2}) are satisfied.

Once the values of $\beta(\tilde{r})$ and $\tilde{p}(\tilde{r})$ are
known, they may be used to obtain the other physical quantities of interest.
In particular,
\begin{eqnarray}
n(\tilde{r})&=&\left [{4p_1\over (1+\sigma_w)\beta_w\gamma_w
m_ec^2}\right ] {1\over
\gamma(\tilde{r})\beta(\tilde{r})\tilde{r}^2}\ , \label{n_e} \\
\frac{\gamma_e(\tilde{r})}{\gamma_w}&=&\left
({3\sigma_w\beta_w\over 2}\right )\tilde{r}^2\gamma(\tilde{r})
\beta(\tilde{r})\tilde{p}(\tilde{r})\ . \label{g_e}
\end{eqnarray}
The postshock value of $n$ is given from equations (\ref{beta_ps}
and (\ref{particle3}) and can, in turn,
be used with equations (\ref{p_e}) and (\ref{p_ps}) to yield the
postshock value of $\gamma_e$,
\begin{equation}\label{gamma_eps}
\frac{\gamma_{e,{\rm ps}}}{\gamma_w} = {3\beta_w \left\{2\gamma_{\rm ps}^2\beta_{\rm
ps}^2-\sigma_w\left[1-(\gamma_{\rm ps}\beta_{\rm ps}/\gamma_w\beta_w)^2\right]\right\}
\over 2(4\gamma_{\rm ps}^2\beta_{\rm ps}^2+1)\gamma_{\rm
ps}\beta_{ps}}\ .
\end{equation}
Furthermore, $\tilde{B}(\tilde{r})$ is given by $1/\tilde{r}\beta(\tilde{r})$
in the ideal-MHD zone and by equation (\ref{B_eq1}) in the equipartition
region. We will also find it useful to consider the variable
\begin{equation}\label{psi}
\psi \equiv \frac{({\mathcal{E^\prime}}+{B^\prime})^2}{8\pi p}\ ,
\end{equation}
which is equal to $p_B/p$ in the ideal-MHD
regime. Note that the ideal-MHD sector of the PWB corresponds to the
region where $\psi(\tilde{r})$ is $< 1$ and that, within this
sector, $\psi=\delta$.

\begin{figure}
\centering
\noindent
\includegraphics[width=13cm]{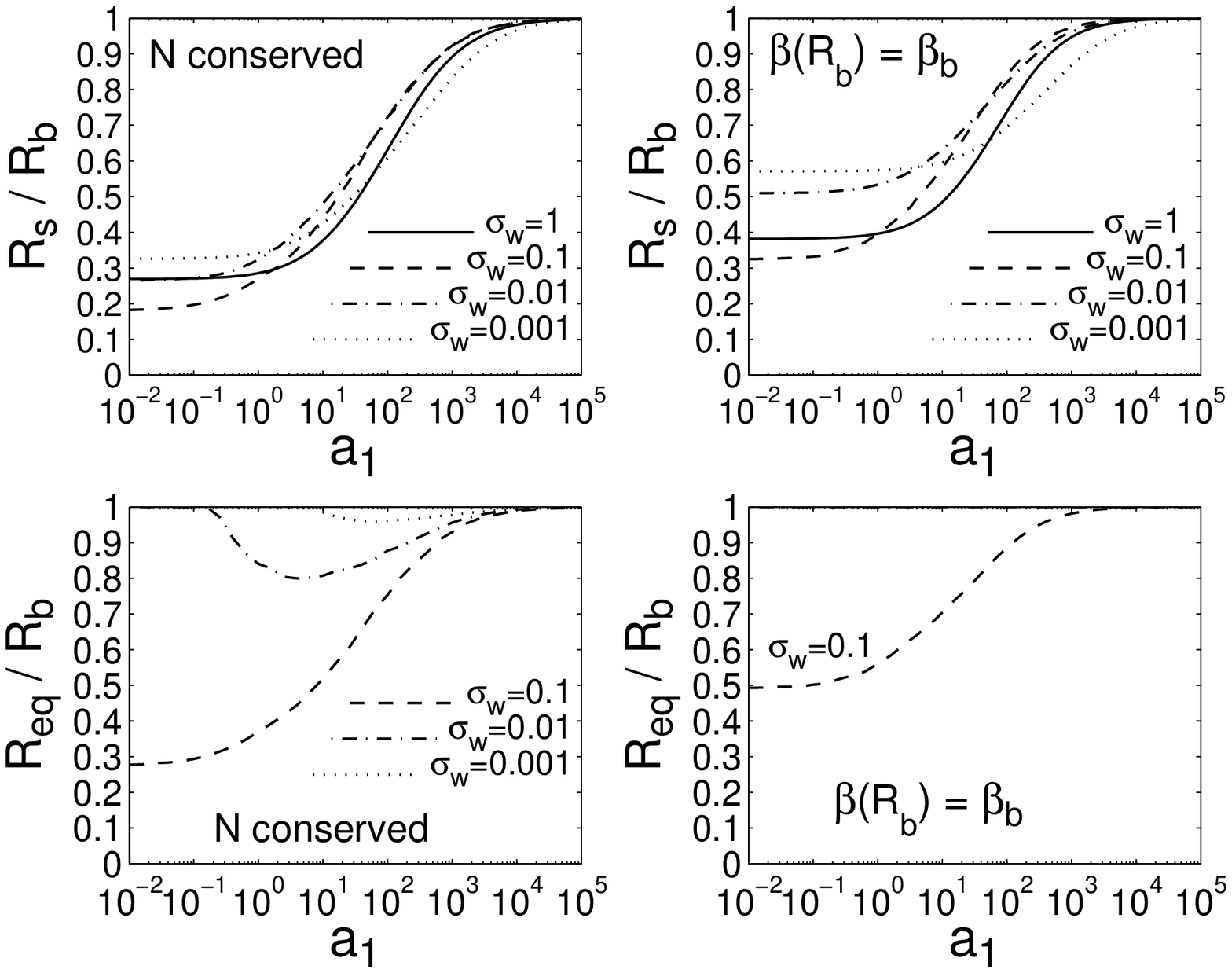}
\caption{\label{fig1} 
The wind-shock radius $R_s$ ({\it top}\/ panels) and the
equipartition radius $R_{\rm eq}$ ({\it bottom}\/ panels), normalized by the
bubble radius $R_b$, as functions of the
parameter $a_1$ (defined in eq. [\ref{a}]) for 4 values of the
pulsar-wind magnetization parameter $\sigma_w$. The {\it left}\/ panels
present solutions obtained under the particle-conservation
constraint (eq. [\ref{BC1}]), whereas the {\it right}\/ panels show
solutions derived by fixing the gas speed at $R_b$
(eq. [\ref{BC2}]).}
\end{figure}

\begin{figure}
\centering
\noindent
\includegraphics[width=13cm]{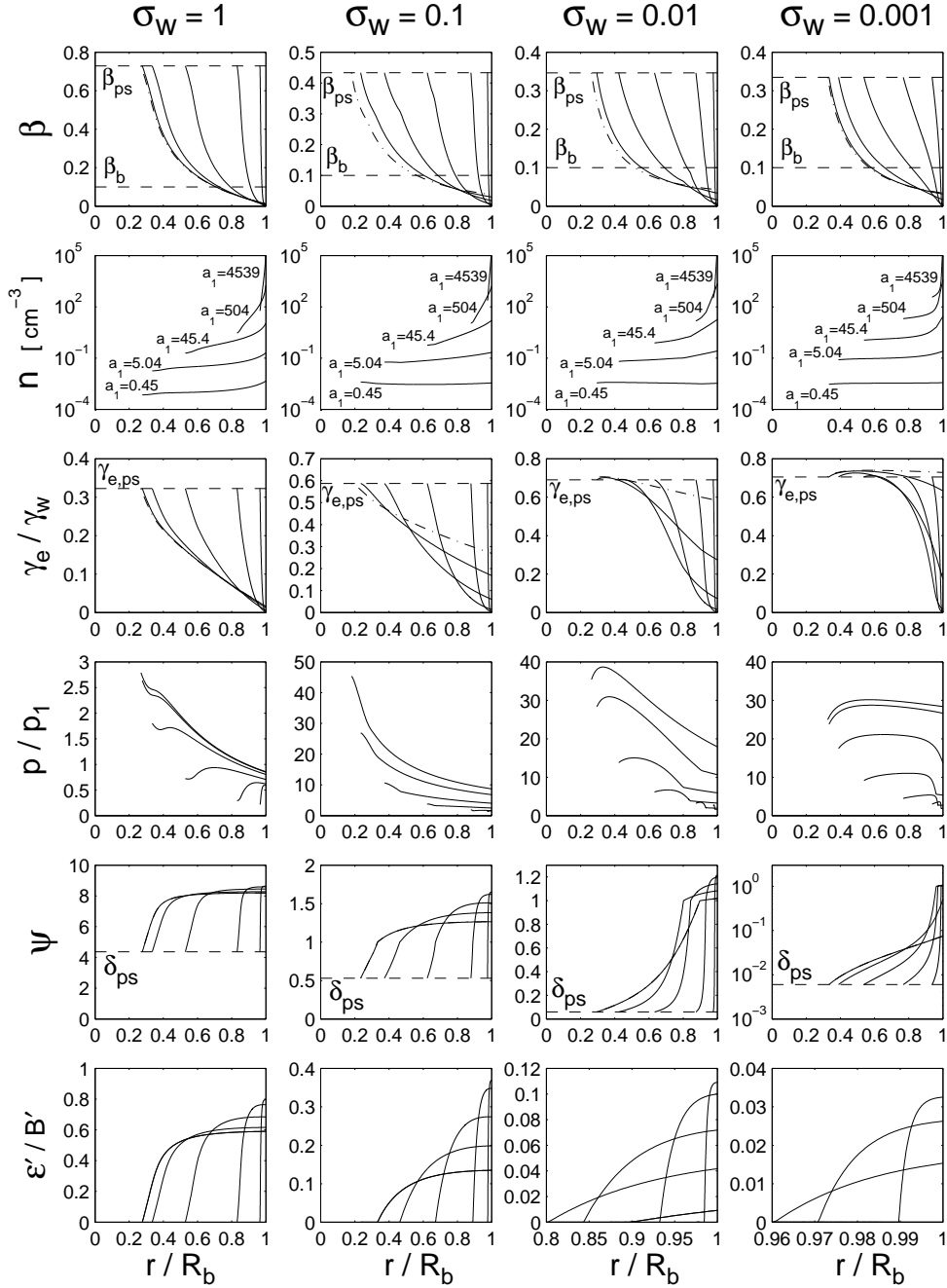}
\caption{\label{fig2} 
The panels show the radial distributions of $\beta$ (the gas speed in
units of the speed of light), $n$ (the particle number density, for which the
dimensional scaling corresponds to the fiducial values of the
model parameters), $\gamma_e$ (the random electron/positron
Lorentz factor, normalized by the pulsar-wind Lorentz factor
$\gamma_w$), $p$ (the particle pressure, normalized by
$p_1$; see eq. [\ref{p_ref}]), $\psi\equiv
(1+{\mathcal{E}}^\prime/B^\prime)^2{B^\prime}^2/8\pi p$, and
${\mathcal{E}}^\prime/B^\prime$ (the ratio of the comoving
electric and magnetic fields) in the model PWBs for 4 values of the
pulsar-wind magnetization parameter $\sigma_w$ (listed at the top
of the respective columns of panels). The solution curves in each
panel correspond to 5 values of the parameter $a_1$: 0.45, 5.04,
45.3, 504, and 4539 (for endpoints progressing respectively from
left to right); they were obtained by imposing the
particle-conservation constraint (eq. [\ref{BC1}]).}
\end{figure}

\begin{figure}
\centering
\noindent
\includegraphics[width=13cm]{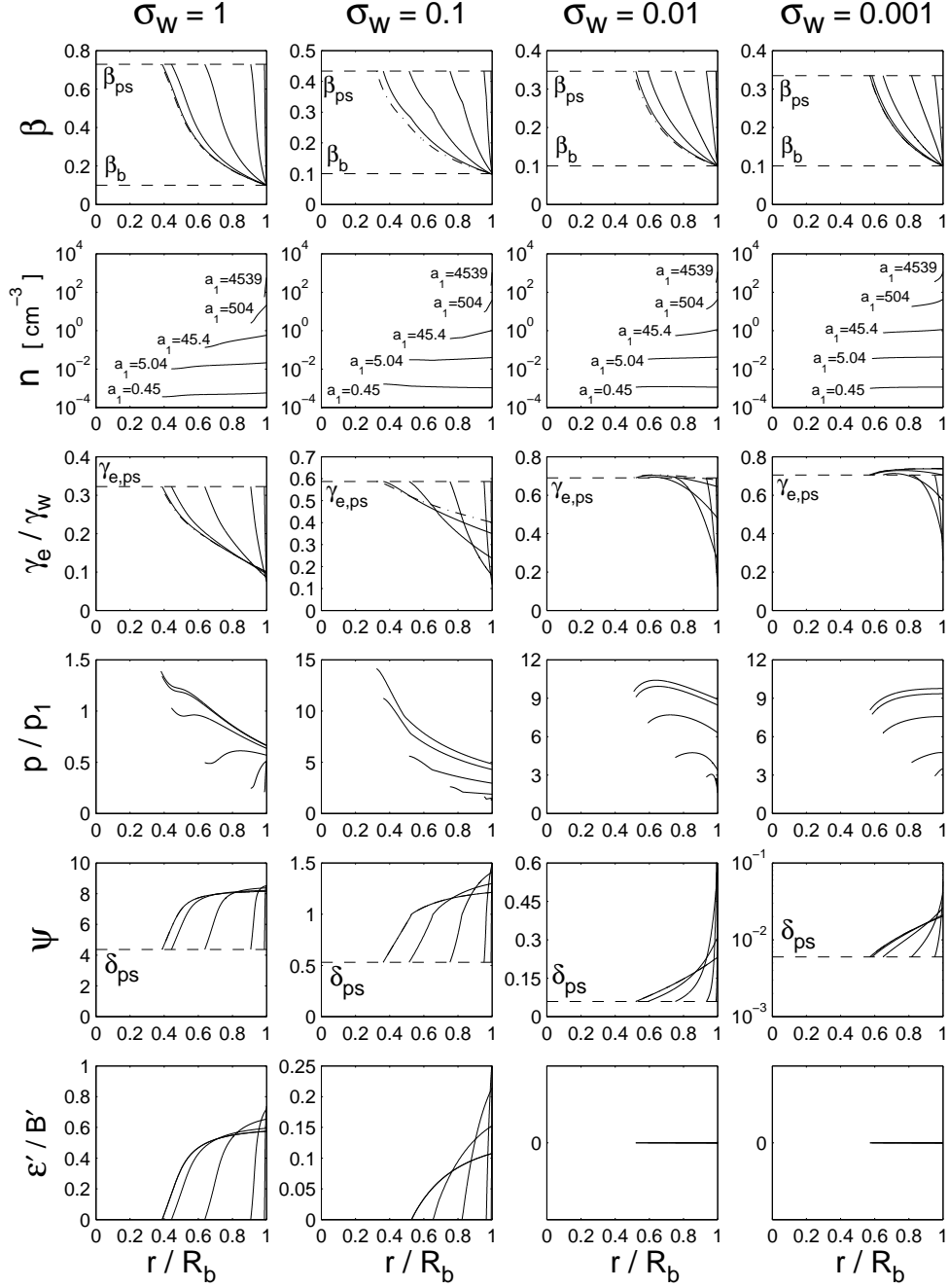}
\caption{\label{fig3} 
Same as Fig. \ref{fig2}, except that the solutions were obtained by using
the final-speed constraint (eq. [\ref{BC2}]) instead of the
particle-conservation constraint.}
\end{figure}

Figure \ref{fig1} shows $\tilde{r}_s=R_s/R_b$, the ratio of the
wind-shock radius to the outer bubble radius, as a function of
the cooling parameter $a_1$ for the two
alternative constraints ($N$ conservation and $\beta(R_b)=\beta_b$)
discussed in \S \ref{structure}. Results are shown for
$\sigma_w=1,\, 0.1,\, 10^{-2}$, and $10^{-3}$. 
According to our adopted scalings, $R_b \propto \tau_{\rm sd}
\propto a_1^{-1/2}$, so $R_b$ decreases with increasing
$a_1$. It is, however, seen that when cooling becomes important,
the relative width of the bubble, $1-\tilde{r}_s$, also
decreases with increasing $a_1$. This can be attributed to the
decrease in the internal thermal pressure brought about by the cooling: a lower
pressure, in turn, requires a shorter length scale to
achieve the needed pressure gradient for decelerating the flow.
For given values of $a_1$ and $\sigma_w$, $\tilde{r}_s$ is
larger when the terminal speed is fixed than when particle
conservation is enforced. This can be attributed to the fact that,
in the former case, the gas speed between $r=R_s$
and $r=R_b$ decreases from $\beta_{\rm ps}$ (eq. [\ref{beta_ps}])
to $\beta_b$, whereas in the latter case it decreases from
$\beta_{\rm ps}$ to $\beta(R_b)<\beta_b$ (see discussion at the
end of \S \ref{structure}): the larger velocity
decrement evidently requires a longer deceleration length $(R_b-R_s)$.
Figure \ref{fig1} also depicts the $a_1$ dependence of
$\tilde{r}_{\rm eq}=R_{\rm eq}/R_b$, the normalized
equipartition radius (where $p_B/p$ first increases to 1
if $\delta_{\rm ps}<1$). It shows that, whereas the $\sigma_w=1$
solution obeys nonideal MHD throughout the shocked-wind bubble
(since $\delta_{\rm ps}>1$ in this case), the nonideal regime
becomes progressively smaller with decreasing $\sigma_w$. This is
a direct consequence of equation (\ref{delta_ps}), which
indicates that $\delta_{\rm ps}$ scales approximately linearly
with $\sigma_w$: the lower the value of
$\delta_{\rm ps}$, the longer it will take for $p_B/p$ in the
postshock flow to rise above 1 (our condition for the
termination of the ideal-MHD regime). The relative extent of the
ideal-MHD region is larger for the fixed-$\beta(R_b)$ solutions,
so much so that the solutions of this type with the two lowest values of
$\sigma_w$ contain no nonideal-MHD zone. This can be understood
from the systematically higher values of $\beta(\tilde{r})$
(see Figs. \ref{fig2} and \ref{fig3}) and
correspondingly lower values of $\tilde{B}(\tilde{r})
\propto 1/\tilde{r}\beta(\tilde{r})$ and hence of $p_B(\tilde{r})/p(\tilde{r})$ in the
fixed-$\beta(R_b)$ solutions in comparison with their
$N$-conserving counterparts.

Figures \ref{fig2} and \ref{fig3} exhibit the $\tilde{r}$ dependence of various
quantities of interest in PWB solutions obtained by imposing
the $N$-conservation and fixed-$\beta(R_b)$ constraints,
respectively, for the same 4 values of the wind magnetization parameter as in
Figure \ref{fig1} (spanning 3 orders of magnitude in $\sigma_w$)
and for 5 values of the cooling parameter (spanning 4 orders of
magnitude in $a_1$). If $\beta_{b,-1}$ and $\Delta E_{53}$ are
both set equal to 1, then the chosen values of $a_1$ ($=0.45,\,
5.04,\, 45.3,\, 504$, and 4539) correspond
to SMNS spin-down times $t_{\rm sd}=100,\, 30,\, 10,\, 3$ and $1\ {\rm yr}$,
respectively. Bubbles with $a_1$ near the
lower end of this range resemble adiabatic PWBs, whereas
configurations with $a_1$ near the upper range are 
highly radiative. Radio pulsars have inferred surface magnetic fields in 
the range $\sim 10^{12}-10^{13}\ {\rm G}$, so, by equation (\ref{t_sd}), a
variation of roughly two orders of magnitude in the value of $t_{\rm sd}$ is
naturally expected.

The upper panels in Figures \ref{fig2} and \ref{fig3} show
the behavior of $\beta(\tilde{r})$. The {\it dashed}\/ lines in
these panels indicate the postshock ($\beta_{\rm ps}$) and
outer-boundary ($\beta_b$) speeds, whereas the {\it
dash-dotted}\/ curves depict the purely adiabatic ($a_1=0$) solutions.
The displayed results confirm that $\beta(R_b)<\beta_b$ in the
$N$-conserving solutions. The greatest discrepancy between
$\beta(R_b)$ and the actual speed ($\beta_b$) of the outer
boundary occurs for $\sigma_w=1$, in which case
$\beta(R_b)/\beta_b$ decreases from $\sim
0.13$ to $\sim 0.0063$ as $a_1$ increases from 0.1 to $10^3$.
We consider this to be a tolerable discrepancy, given (as we
already noted in \S \ref{structure}) that the flow near
$\tilde{r}=1$ is highly subsonic.
The $\beta(\tilde{r})$ curves further demonstrate that the values of
$\beta$ in the $N$-conserving solutions are lower than those in
the corresponding
$\beta(R_b)=\beta_b$ solutions also for all other values of
$\tilde{r}$ where the respective solutions overlap. The lower
values of $\beta$ lead (on account of the particle flux-conservation
relation [\ref{particle}]) to systematically higher values of $n$
in the $N$-conserving solutions (see the second row of panels in
Figs. \ref{fig2} and \ref{fig3}). The radial profiles of $n$ are
nearly flat for low values of the cooling parameter, but when
radiative effects are important and contribute to the compression,
$n(\tilde{r})$ rises monotonically between $\tilde{r}_s$ and
$\tilde{r}=1$ with a slope that is steeper (particularly in the
nonideal-MHD regime) the larger the value of $a_1$.

The third row of panels in Figures \ref{fig2} and \ref{fig3}
displays $\gamma_e(\tilde{r})/\gamma_w$. The {\it dashed}\/ line
marks the postshock value of this quantity, and the {\it
dash-dotted}\/ curves again represent the purely adiabatic case.
The random Lorentz factor declines monotonically with
$\tilde{r}$ in the $\sigma_w=1$ and 0.1 solutions, but at
lower values of $\sigma_w$ (for which the
electromagnetic pressure contribution is small) and so long as
cooling remains relatively unimportant ($a_1\lesssim 10$),
$\gamma_e(\tilde{r})$ initially increases behind the wind shock to make it
possible to attain the necessary total pressure. 
The minimum value of $\gamma_e(\tilde{r})$
(reached at $\tilde{r}=1$) decreases (as expected) with increasing $a_1$ and
is lower for the $N$-conserving solutions than for the
fixed-$\beta(R_b)$ ones. This difference can be understood from the fact
that the particle density near the outer boundary is lower in
the latter case [corresponding to a higher value of
$\beta(\tilde{r}=1)$], so $\gamma_e$ must be
larger to bring the pressure up to its requisite value.

Of particular relevance to the
evolution of GRB afterglows is the behavior of the thermal and
electromagnetic pressures. The radial profiles of $p$ (normalized by
$p_1$) are shown in the fourth row of panels in Figures \ref{fig2}
and \ref{fig3}. It is seen that the behavior of $p(\tilde{r})$ varies
with the parameter choices and that the details depend on the
nature of the constraint imposed on the solution. For
$\sigma_w=1$, the curves decline monotonically when cooling is
relatively unimportant but increase monotonically at high
values of $a_1$. For $\sigma_w=0.1$, the curves decrease
monotonically for all plotted values of the cooling parametr, whereas for
$\sigma_w=10^{-3}$ in the fixed-$\beta(R_b)$ solution they increase
monotonically for all exhibited values of $a_1$. In the
remaining cases, the curves increase with $\tilde{r}$ near the
inner boundary of the bubble and decrease near its outer
boundary. Since $p \propto \gamma_e n$, one can
understand the shape of the curves by comparing them with the
corresponding curves in the second and third rows of panels.
One finds that, quite generally, the behavior of $p$ is
dominated by that of $\gamma_e$ for low values of $a_1$, but
that the influence of the density variations becomes
progressively more important as the cooling parameter increases.

The effect of the electromagnetic fields can be inferred from the
behavior of the variable $\psi$ and of the ratio
${\mathcal{E}}^\prime/B^\prime$, shown in the
bottom two rows of panels in Figures \ref{fig2} and \ref{fig3}.
Note that one can write $\psi = \delta +
{\mathcal{E}}^\prime B^\prime/4\pi p$ (see eq. [\ref{psi}]). Immediately behind the
wind shock (in which, by construction, ${\mathcal{E}}^\prime=0$),
$\psi$ is equal to $\delta_{\rm ps}$ (given by
eq. [\ref{delta_ps}]). If $\delta_{\rm ps}\ge 1$ then the
nonideal-MHD regime starts right there: beyond that point, $\delta$ remains fixed
at its postshock value but $\psi$ and
${\mathcal{E}}^\prime/B^\prime$ increase with $\tilde{r}$. If
$\delta_{\rm ps}< 1$, then the evolution proceeds under
ideal-MHD conditions (${\mathcal{E}}^\prime=0$), with $\psi$
continuing to coincide with $\delta$ (which in this regime equals
$p_B/p$) up to the point where it reaches 1. This point
marks the end of the ideal-MHD regime 
and corresponds to $\tilde{r}_{\rm eq}$. Beyond $\tilde{r}_{\rm
eq}$, $\psi$ continues to increase monotonically
with $\tilde{r}$ (and now so does also
${\mathcal{E}}^\prime/B^\prime$), but $\delta$ remains fixed at 1 (see
eq. [\ref{delta}]). The figures show that, as expected, the
electromagnetic contribution to the total pressure becomes
progressively larger as $\sigma_w$ increases, and they
further demonstrate that ${\mathcal{E}}^\prime/B^\prime$ exhibits a
similar trend. It is also seen that cooling enhances the
relative importance of the electromagnetic fields, which can be
understood from the fact that it reduces the magnitude of the
thermal pressure component. The overall behavior of $\psi$ and
${\mathcal{E}}^\prime/B^\prime$ does not appear to depend
strongly on the choice of constraint under which the solution is
obtained, although the corresponding curves in the two figures differ in their
details (see also the related discussion in connection with
Fig. \ref{fig1} above). As we show in \S \ref{parameters}, the
variable $\psi$ plays a key role in the modeling of relativistic shocks that
propagate inside a PWB.

\section{Implications to GRB Afterglows}
\label{discussion}
\subsection{Emission-Region Parameters}
\label{parameters}

Pulsar wind-inflated bubbles, such as those predicted to arise prior to the 
onset of the high-energy burst in the supranova scenario, provide an optimal 
environment for GRB afterglows since they naturally yield high electron and
magnetic energy fractions ($\epsilon_e$ and $\epsilon_B$) behind
the propagating shock wave that gives rise to the afterglow
emission. High values
of $\epsilon_e$ are expected from the fact that relativistic pulsar-type winds
are likely dominated by an electron-positron component, whereas
significant values of $\epsilon_B$ should naturally occur if the
winds are characterized by a high magnetization parameter.

The observationally inferred values of $\epsilon_e$ and $\epsilon_B$ are 
derived from spectral fits that are based on the standard model
assumptions of a ``cold,'' weakly magnetized proton-electron
preshock medium. We now consider what would be the
``equivalent'' values (which we
denote by the subscript \textsf{equiv}) that one would derive if
the afterglow-emitting shock propagated instead inside a PWB. 
The postshock quantities can be determined from the appropriate
generalizations of the expressions presented in \S
\ref{structure}, taking account of the fact that the preshock
gas is now ``hot'' (with $w=4p$; eq. [\ref{rel_eos}]) rather
than ``cold'' and that it may contain a nonzero comoving electric field. 
One imposes the continuity of the energy flux $\gamma^2\beta
w+{\mathcal{E}}B/4\pi$, momentum flux $\gamma^2\beta^2w +p
+({\mathcal{E}}^2+B^2)/8\pi$, and tangential electric field
${\mathcal{E}}$ in the frame of the shock. The possible presence
of a nonzero ${\mathcal{E}}^\prime$ requires the specification
of an additional shock jump condition, which we take to be the
conservation of magnetic flux during the fluid's transit
through the shock. On the assumption that the field is
transverse to the shock propagation direction, this condition
translates into the requirement that $\beta B$ be continuous in
the shock frame. Combining these constraints, one obtains a cubic equation
for the postshock (subscript 2) flow speed (measured in the shock
frame) that is akin to the expression (\ref {eq_beta_ps}). In
the limit of an ultrarelativistic shock, the latter again
reduces to a quadratic equation, whose relevant solution is
\begin{equation}\label{beta_2}
\beta_2 = {1+\psi+[(1+\psi)^2+3(2+\psi)\psi]^{1/2} \over
3(2+\psi)}\ ,
\end{equation}
where $\psi$ is defined by equation (\ref{psi}). Equation
(\ref{beta_2}) reproduces equation (\ref{beta_ps}) if one
substitutes $2\sigma_w$ for $\psi$: this is
consistent with the fact that $\psi/2$ becomes equal to the
magnetization parameter $\sigma \equiv
{\mathcal{E}}B/4\pi\gamma^2\beta w$ of the preshock medium in the
limit $\beta \rightarrow 1$ and $w\rightarrow 4p$.\footnote{An
indirect measure of the value of the magnetization parameter
just ahead of the afterglow-producing shock is
possibly provided by the power-law index $p$ of the
synchrotron-emitting particle energy distribution, which can be deduced from 
the shape of the observed spectrum (e.g., \cite{SPN98}). Kirk et al. (2000) 
argued that an ultrarelativistic, unmagnetized shock that accelerates 
particles in a ``cold'' medium through the first-order Fermi
process produces a ``universal'' power law of index $p\approx
2.2$, and that this value increases with the preshock
magnetization (so that, for example, $p\approx 2.3$ for 
$\sigma = 0.01$, assuming ${{\mathcal{E}}^\prime}=0$). The
often-quoted ``canonical'' value of $p$ for GRB 
afterglows is 2.5, although in some sources a significantly
higher value has been inferred (e.g., \cite{HDL00};
\cite{PK02}---note, however, that the latter reference also
lists a few sources in which $p<2.2$ has been deduced).}

\begin{figure}
\centering
\noindent
\includegraphics[width=13cm]{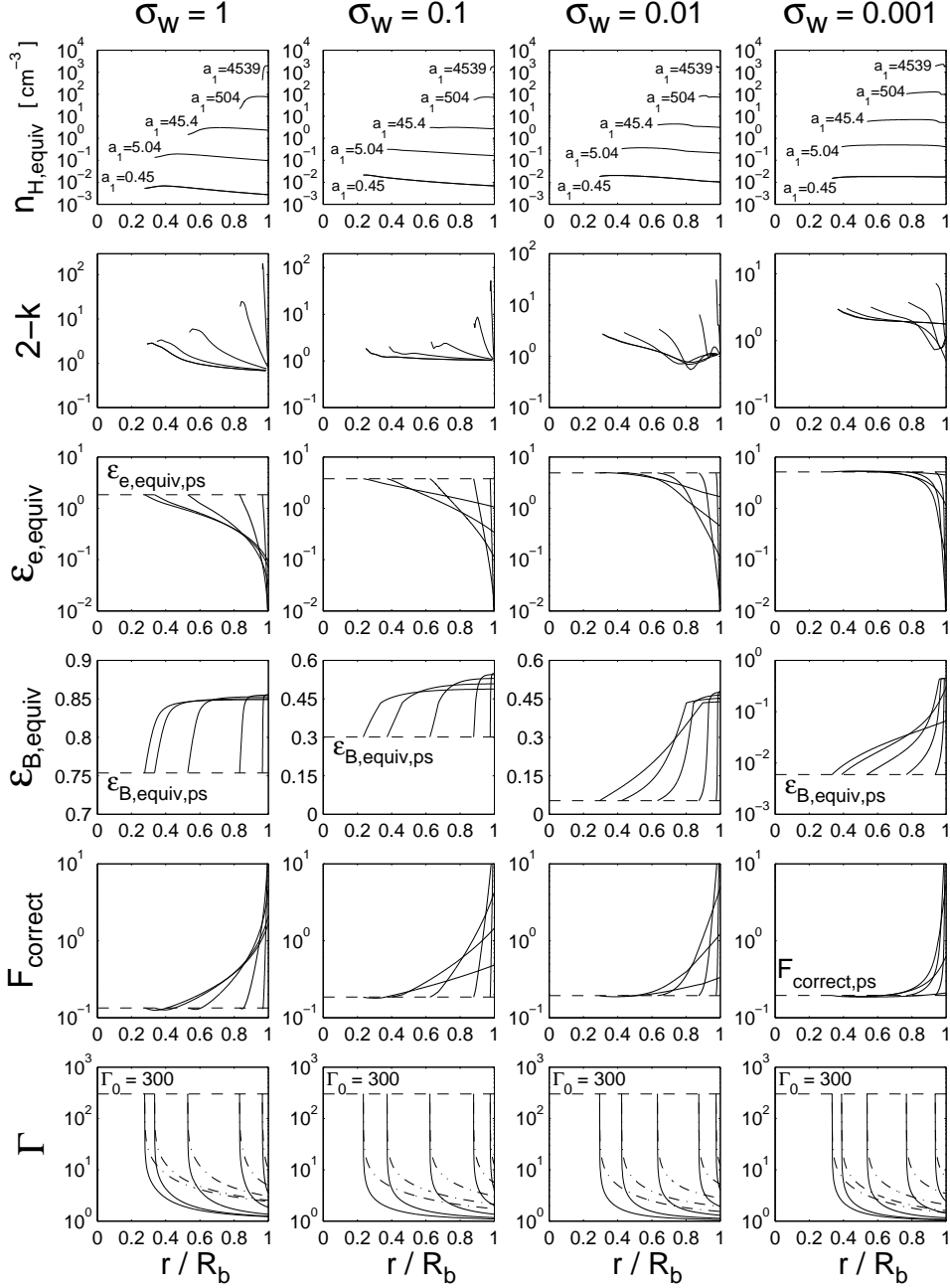}
\caption{\label{fig4} 
The panels show the radial distributions of $n_{\rm H, equiv}$
(the effective hydrogen number density, eq. [\ref{n_eq}]), $2-k$
(with $k\equiv -d\log{n_{\rm H, equiv}} /d\log{r}$ being the
effective power-law 
index of the equivalent hydrogen density distribution),
$\epsilon_{e,{\rm equiv}}$ (the equivalent electron energy
fraction, eq. [\ref{epsilon_e_eq}]), $\epsilon_{B,{\rm equiv}}$
(the equivalent magnetic energy fraction,
eq. [\ref{epsilon_B_eq}]), $F_{\rm correct}=n/n_{\rm H, equiv}$
(the flux correction factor, eq. [\ref{F_correct}]), and the
Lorentz factor of the shocked bubble
material (for a spherical shock driven into the PWB by an outflowing mass
of energy $E$ and initial Lorentz factor $\Gamma_0$) for the PWB
solutions depicted in Fig. \ref{fig2}. The displayed results
correspond to the fiducial values of the model parameters.
The {\it solid}\/ and {\it dash-dotted}\/ curves in the bottom panels 
correspond to $E=10^{52}$ and $10^{53}\ {\rm ergs}$, respectively.}
\end{figure}

\begin{figure}
\centering
\noindent
\includegraphics[width=13cm]{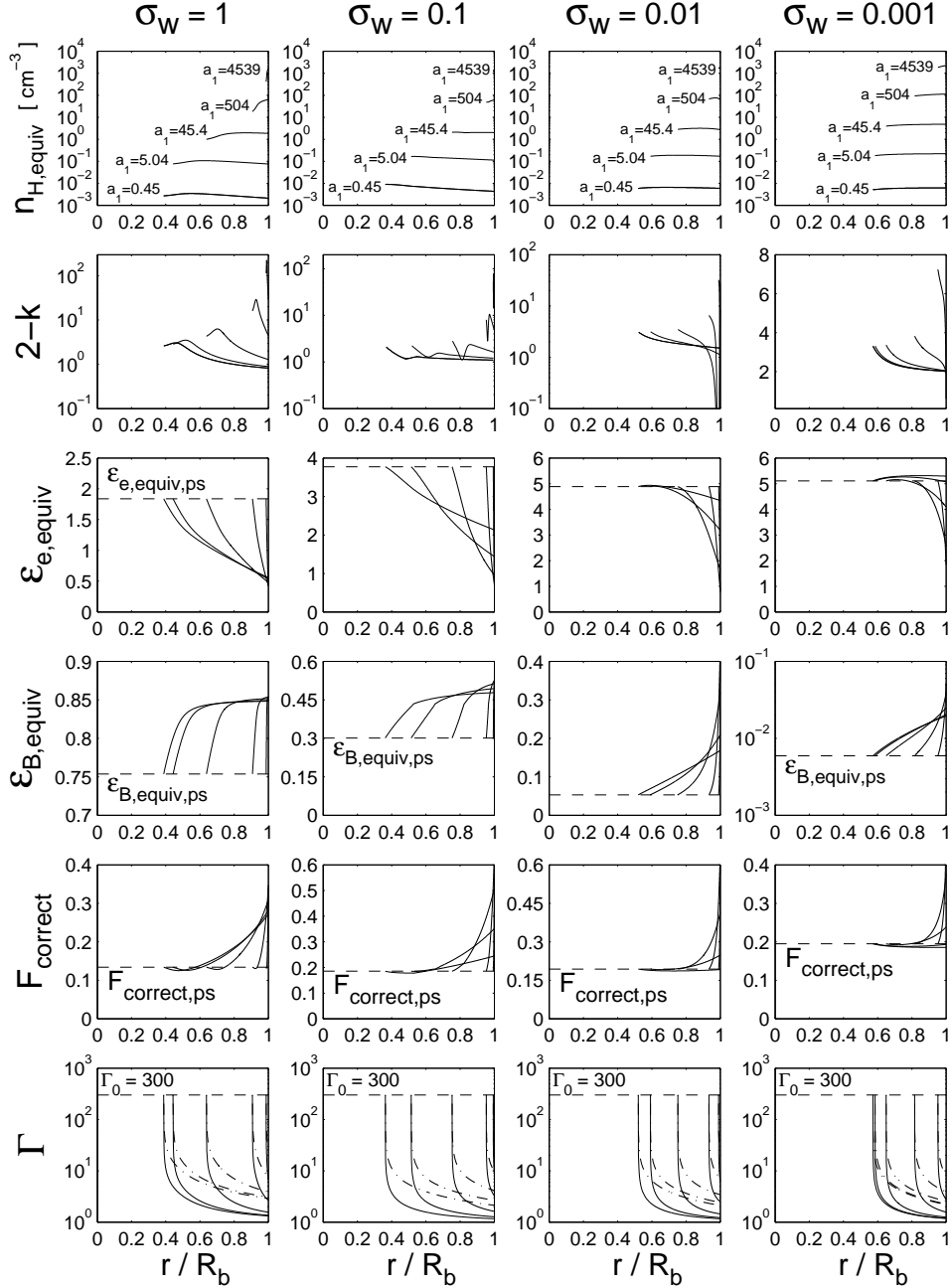}
\caption{\label{fig5} 
Same as Fig. \ref{fig4}, except that the results correspond to
the PWB solutions depicted in Fig. \ref{fig3}.}
\end{figure}

By substituting ${\mathcal{E}}=\gamma({\mathcal{E}}^\prime +
\beta B^\prime)$ and $B = \gamma (B^\prime +
\beta{\mathcal{E}}^\prime)$ (see eq. [\ref{EB}]) into the
expressions for the energy and momentum fluxes in the shock frame, one
finds that, if the shock is highly relativistic (so that the
upstream $\beta$ is close to 1 in the frame of the shock), the upstream fluxes
have the same forms as in a purely hydrodynamic shock if $w$ in
the latter is replaced by
$w+(B^\prime+{\mathcal{E}}^\prime)^2/4\pi$. Shock models of GRB
afterglows traditionally infer an ambient gas density by
assuming that a hydrodynamic shock propagates into a standard ISM or
stellar-wind environment with an enthalpy density $w= n_{\rm H} m_p c^2$, where
$m_p$ is the proton mass.
This motivates us to define the ``equivalent'' hydrogen number density
\begin{equation}\label{n_eq}
n_{\rm H, equiv} \equiv \frac{1}{m_p c^2}\left [w +
\frac{(B^\prime+{\mathcal{E}}^\prime)^2}{4\pi}\right ]\ ,
\end{equation}
which under the assumption of a relativistic equation of state
can be written as $n_{\rm H, equiv}=4(1+\psi/2)p/m_pc^2$. This
quantity is plotted in the top row of panels in Figures
\ref{fig4} and \ref{fig5} as a function of $\tilde{r}$ for each
of the model PWBs presented in \S \ref{results}.
These figures also show (in the second row of panels) plots of the
radial dependence of
$k\equiv -d\log{n_{\rm H, equiv}} /d\log{r}$, the effective power-law 
index of the equivalent hydrogen density distribution. Our model
assumption of an abrupt transition between the ideal- and
nonideal-MHD regimes at $\tilde{r}_{\rm eq}$ introduces an
unphysical discontinuity in the value of $k$ at this point: the
displayed curves have been smoothed at this
location by interpolation across the discontinuity.

To simplify the discussion, we restrict attention
to the three synchrotron-spectrum characteristics considered by
Sari et al. (1998, hereafter SPN), namely, the break frequencies $\nu_m$ and
$\nu_c$ and the peak flux $F_{\nu,{\rm max}}$; we refer the
reader to that paper for the definition of these quantities and
for the derivation of the standard expressions for emission by a
spherical shock in which there is no pair production. In the
interest of simplicity, we also assume that the equivalent hydrogen number
density inside the bubble is roughly constant; as the $n_{\rm
H,equiv}$ plots in Figures \ref{fig4} and \ref{fig5}
demonstrate, this approximation is
usually adequate over the bulk of the bubble volume, especially
when the cooling is not too strong.
We distinguish between two cases: weakly cooling PWBs
(corresponding to cooling parameters $a_1\lesssim 10^2$, or, for
our fiducial values, $\tau_{\rm sd}\gtrsim 10$), whose
radial widths $\Delta R_b \equiv (R_b-R_s)$ are of the order of $R_b$,
and strongly cooling PWBs ($a_1\gg 10^2$, $\tau_{\rm sd} \ll 10$),
for which $\Delta R_b/R_b \ll 1$. In the weakly cooling case, one
can approximate the volume of the shocked bubble gas by that of the
sphere that is bounded by the shock.

In the standard case of a
uniform ambient medium and a slow-cooling (adiabatic) shock, one
can express the two break frequencies and the peak flux in terms of the
shock energy $E$, the ambient density $n_{\rm H}$, the observed time
$t$, as well as $\epsilon_e$, $\epsilon_B$, and the
distance to the source (see eq. [11] in SPN). In particular, $\nu_m \propto
\epsilon_e^2\epsilon_B^{1/2}E^{1/2}t^{-3/2}$, $\nu_c
\propto \epsilon_B^{-3/2}E^{-1/2}n_{\rm H}^{-1}t^{-1/2}$, and $F_{\nu,{\rm
max}} \propto \epsilon_B^{1/2}En_{\rm H}^{1/2}$.  In the case
of a shock propagating inside a weakly cooling PWB, it turns out
that the above expressions for $\nu_m$ and $\nu_c$
are reproduced if $n_{\rm H}$, $\epsilon_e$, and
$\epsilon_B$ are everywhere replaced by $n_{\rm H, equiv}$ (eq. [\ref{n_eq}]),
\begin{equation}\label{epsilon_e_eq}
\epsilon_{e,{\rm equiv}} \equiv \frac{m_e}{m_p}
\frac{\gamma_{e2}}{\gamma_{21}}\epsilon_e = \left ( \frac{m_e}{m_p} \right )
\left (\frac{1+\beta_2}{\beta_2}\right ) \left (
\frac{4\gamma_2^2\beta_2^2 - \psi}{4\gamma_2^2\beta_2^2+1}
\right ) \epsilon_e \gamma_e\ ,
\end{equation}
and
\begin{equation}\label{epsilon_B_eq}
\epsilon_{B,{\rm equiv}} \equiv
\frac{{B_2^\prime}^2}{32\pi\gamma_{21}^2n_{\rm H,equiv}m_pc^2}= \left (
\frac{1+\beta_2}{4\beta_2}\right )^2
\left (\frac{1}{2} + \frac{1}{\psi}\right )^{-1}\ ,
\end{equation}
respectively, 
where $\gamma_2$ is the Lorentz factor corresponding to
$\beta_2$, $\gamma_{21}$ is the Lorentz factor of the postshock
fluid in the stationary frame (evaluated again on the assumption that
the afterglow-emitting shock is highly relativistic), and
$\gamma_e$ is given by equation (\ref{g_e}).
In deriving equation (\ref{epsilon_e_eq}), we have made use of
equation (\ref{p_e}) and of the continuity of the particle flux
$\gamma \beta n$ in the shock frame. The expression for $F_{\nu,{\rm
max}}$ is reproduced by making similar substitutions and then
multiplying by the factor
\begin{equation}\label{F_correct}
F_{\rm correct} =  \frac{n}{n_{\rm H,equiv}} = \frac{3}{4}
\frac{m_p}{m_e}\left (1+\frac{\psi}{2}\right )^{-1} \gamma_e^{-1}\ .
\end{equation}
The functions $\epsilon_{e,{\rm equiv}}(\tilde{r})$,
$\epsilon_{B,{\rm equiv}}(\tilde{r})$, and $F_{\rm
correct}(\tilde{r})$ are plotted in the third-through-fifth rows,
respectively, of Figures \ref{fig4} and \ref{fig5}, with the
{\it dashed}\/ line in each panel marking the value of the
respective quantity immediately behind the pulsar-wind
shock. Although the underlying expressions were obtained under
the assumption of a weakly cooling bubble, it turns out (see
next paragraph) that they continue to apply also in the strongly
cooling case. The values of $\epsilon_{e,{\rm equiv}}$ were
calculated by setting
$\epsilon_e=1$ in equation (\ref{epsilon_e_eq})--- consistent
with our PWB model approximation of a pure-$e^\pm$ wind.
Inasmuch as $F_{\rm correct}$ does not differ from 1 by
more than a factor of a few over most of the explored parameter
range, it can be concluded that the standard expressions will
remain approximately applicable if one simply replaces $n_{\rm
H}$, $\epsilon_e$, and $\epsilon_B$ by their ``equivalent''
counterparts. The derived values of $F_{\rm correct}$ also
verify that $n_{\rm H, equiv}$ is usually much greater than
$(m_e/m_p)n$, consistent with the PWB model assumption of a
``hot'' equation of state.

To derive the corresponding expressions for a
strongly cooling PWB, one can approximate the bubble as a thin shell
of radius $\sim R_b$. The bubble volume traversed by a shock
that is located at a distance $x$ from the inner radius of the bubble is
then $V(x)\approx 4\pi R_b^2x$. Relating $x$ to the
observed time $t$
and to the Lorentz factor $\Gamma$ of the shocked gas by
$x\approx 4 \Gamma^2 t$, following SPN,\footnote{In a more precise
treatment, one obtains $t$ for radiation emitted along the line
of sight to the center from the differential equation $c\, dt/dr
= 1/2 \Gamma_{\rm sh}^2$, where $\Gamma_{\rm sh}$ is the Lorentz factor of the
shock (see \cite{S97}). The solution in this
case is $t=x/4\Gamma_{\rm sh}^2c+R_s/2\Gamma_0^2c$ (where
$\Gamma_0$ is the initial Lorentz factor of the outflow and
$\Gamma_{\rm sh}^2 \approx
[(1+\beta_2)/(1-\beta_2)]\Gamma$ for $\Gamma_{\rm sh}\gg1$), which shows that
the approximation used in the text is only valid for $x\gg
(\Gamma/\Gamma_0)^2R_s$.} and setting
$E\approx V
\Gamma^2 n_{\rm H,equiv} m_p c^2$ in the adiabatic-shock case,
one finds that the standard expressions (eq. [11] in
SPN) continue to apply if one makes the aforementioned substitutions
for $\epsilon_e$, $\epsilon_B$, and $n_{\rm H}$, and, in
addition, multiplies the expressions for $\nu_m$ and $\nu_c$ by
$A_{\rm correct}$ and $1/A_{\rm correct}$, respectively, where
\begin{equation}\label{A_correct}
A_{\rm correct} = \left ( \frac{4Et}{17\pi c m_p n_{\rm
H,equiv}R_b^4} \right )^{1/2}\, =\, 3.59\, E_{52}^{1/2}
t_d^{1/2}n_{\rm H,equiv}^{-1/2}R_{b,17}^{-2}\ .
\end{equation}
Here $E_{52}\equiv(E/10^{52}\ {\rm ergs})$, $t_d$ is the observed
time in units of days, and $R_{b,17}\equiv (R_b/10^{17}\ {\rm
cm}$ (see eq. [\ref{R_b}]). The expression for the flux
correction factor remains the same as in the weakly-cooling-PWB
case; the displayed plots indicate, however, that $F_{\nu,{\rm max}}$ 
undergoes large variations across the bubble when the particle-conservation 
constraint is imposed (although not when the terminal
speed is fixed). Although the factor $A_{\rm correct}$ alters the parameter
dependences of the break frequencies (specifically, $\nu_m
\propto E t^{-1} n_{\rm H,equiv}^{-1/2}$ and $\nu_c \propto
E^{-1} t^{-1} n_{\rm H,equiv}^{-1/2}$ for an adiabatic shock in
a strongly cooling bubble), its numerical value will not be large
for typical afterglow parameters. This seems to suggest that the
standard expressions should provide adequate estimates of the
source parameters in this case, too, but we caution that the adopted
approximation of an effectively uniform ambient medium becomes
questionable at large values of $a_1$.\footnote{In the
conventional interpretation, the forward shock is
expected to be fast-cooling (and, if $\epsilon_e$ is close to 1,
also radiative) during the early phase of the afterglow
evolution. However, in the case of a PWB environment
characterized by $\sigma_w \lesssim 1$, the
shock can be only partially radiative since a significant fraction of
the shock kinetic energy is converted into postshock magnetic
energy, which is not subject to radiative losses. Although one
can in principle derive the appropriate expressions also for
this situation [as was done, e.g., by B\"ottcher \& Dermer (2000) in the
standard case], we do not consider fast-cooling shocks here
since the results for partially radiative shocks are
more cumbersome and would unduly complicate the presentation.}

The above considerations suggest that, as a rough check of the
compatibility of the PWB model with observations, one can
examine the consistency of the predicted values of
$\epsilon_{e,{\rm equiv}}$, $\epsilon_{B,{\rm equiv}}$, and
$n_{\rm H,equiv}$ with the values of $\epsilon_e$, $\epsilon_B$,
and $n_{\rm H}$ that are inferred from the spectral data by using
the standard ISM model. The results shown in Figure \ref{fig4}
indicate that, if $\sigma_w$ is not $\ll 1$, then $\epsilon_{e,{\rm
equiv}}$ typically lies in the range $\sim 0.1-1$.
As we noted in \S 1, such comparatively large values have been
inferred for the corresponding standard parameter $\epsilon_e$
even before a large body of data became available, based on
emission-efficiency considerations as well as on some early model fits.
These inferences have been supported by more recent analyses of
the accumulating data on afterglows, which have even led to the
suggestion that $\epsilon_e$ may have a ``universal'' value $\sim 0.3$ 
(\cite{FW01}; see also \cite{HDL00} and \cite{PK02}). The
predicted magnitudes of $\epsilon_{e,{\rm equiv}}$ are larger
for smaller values of $\sigma_w$ and when the fixed-$\beta(R_b)$
constraint is imposed (see Fig. \ref{fig5}). However, these
estimates could be lowered in real sources by the admixture of
baryons into the bubble interior --- either by injection at the source
or by evaporation from the bounding supranova-ejecta shell (see
footnote \ref{evaporation}).
Turning next to the magnetic energy-density parameter
$\epsilon_{B,{\rm equiv}}$, we see from Figures \ref{fig4} and
\ref{fig5} that values in the range $\lesssim 1$ to $\lesssim
10^{-2}$ are predicted by our model as $\sigma_w$ decreases from
1 to $10^{-3}$. It is noteworthy that this range is consistent
with the values inferred in the standard ISM picture for a source like GRB
970508 (which, as we discussed in \S 1, have posed a challenge
for the conventional scenario) as well as with the mean values
of $\epsilon_B$ inferred for the afterglow sample of Panaitescu
\& Kumar (2001). In this case one can again expect a reduction in the estimated
parameter value as a result of the admixture of
baryons (which would reduce the cooling and the associated magnetic field
compression in the bubble; see GK), although other factors may also
contribute to a lowering of $\epsilon_B$: for example, the
afterglow-emitting shock may not be transverse, and the 
electromagnetic-to-thermal pressure ratio in the diffusive
regions of the PWB might be lower than the equipartition
value adopted in our model. 

Finally, our derived values of $n_{\rm H,equiv}$ are compatible
with the observationally inferred preshock particle
densities. The values of $n_{\rm H}$ estimated in
the literature under the assumption of a uniform, ``cold''
ambient medium typically span the range $\sim 0.1-50\ {\rm
cm^{-3}}$ (e.g., \cite{PK02}), although (as noted in the
last-cited reference) there are examples of sources where a
density $< 10^{-2}\ {\rm cm^{-3}}$ is implied. There have also
been suggestions in the literature that some afterglows
originated in a medium with a density $>10^2\ {\rm cm^{-3}}$.
As is seen from Figures \ref{fig4} and \ref{fig5}, our model can in
principle account for all of the inferred values: the
typical densities are reproduced by PWBs with $\tau_{\rm sd}$
in the range $\sim 3-30$, whereas ``outlying'' low and
high inferred densities correspond to lower (respectively, higher)
values of the cooling parameter $a_1$. The basic trend is for bubbles with more
radiative cooling to be characterized by higher values of the
equivalent density:\footnote{In particular,
afterglows with inferred preshock densities
above $\sim 10\ {\rm cm^{-3}}$ are expected in this picture to
arise in highly radiative PWBs, suggesting that such sources may be the most
promising candidates for testing the predicted departures from the
standard spectral scaling relations (which, according to our preceding
arguments, should be most pronounced in rapidly cooling bubbles).}
this follows directly from our adopted parameterization (see \S
\ref{scenario}), which implies $n_{\rm H,equiv} \propto p_1
\propto \tau_{\rm sd}^{-3} \propto a_1^{3/2}$.
It is, furthermore, seen that the effective-density predictions are
remarkably insensitive to the choice of the wind magnetization
parameter $\sigma_w$ and of the specific constraint imposed on the
solution. This robustness can be traced to the fact that $n_{\rm
H,equiv}$ basically measures the energy density in the bubble,
which, for a given choice of $a_1$ (and thus of $R_b\propto
a_1^{-1/2}$), is essentially determined by the wind ram pressure
at a distance $\sim R_b$ from the center.

Another attractive feature of the PWB scenario is that it naturally gives
rise to radial profiles of $n_{\rm H, equiv}$ that, depending on the
cooling parameter $a_1$ and the location within the bubble (see
the plots of $2-k$ in Figs. \ref{fig4} and \ref{fig5}), may
resemble a uniform medium (constant-$n_{\rm
H}$ ISM or interstellar cloud) or a stellar wind ($n_{\rm
H}\propto r^{-2}$; but note that $k$ strictly remains $\gtrsim
1$ in these solutions). Both types of behavior have, in fact,
been inferred in afterglow sources (e.g., \cite{CL00}; 
\cite{F00}; \cite{H00}). The unique aspect of the radial 
distribution of $n_{\rm H, equiv}$ in this picture is that it spans a range of
effective power-law indices $k$ that can vary from source to source, and,
moreover, that the value of $k$ appropriate to any given afterglow is predicted
to change with time as the afterglow-emitting shock propagates within the bubble.
This leads to a more flexible modeling framework for the
afterglow evolution and can naturally
accommodate cases where a value of $k$ that is intermediate between those of
a uniform ISM and a stellar wind could best fit the observations 
(see, e.g., \cite{LW00}). It also explains why afterglows
associated with star-forming regions need not show evidence of
a stellar-wind environment (as expected when the GRB progenitor
is a massive star; in view of the derived magnitudes of $n_{\rm
H, equiv}$, this model also makes it possible to understand how
a source with such a
progenitor could produce an afterglow with an implied value of
$n_{\rm H}$ that was much lower than the typical ambient
density near massive stars). In addition, high values of $n_{\rm H, equiv}$ in
this picture are not subject to the objection (e.g., \cite{H00}) that they 
will necessarily give rise to excess extinction (although it is
also conceivable that dust destruction by the optical-UV
and X-ray radiation from the GRB outflow could reduce
any preexisting extinction toward the source; see \cite{WD00}
and \cite{FKR01}). As is seen from Figures \ref{fig4} and \ref{fig5}, the
predicted $n_{\rm H, equiv}(r)$ distributions exhibit
progressively steeper declines as the outer
boundary of the bubble is approached. This suggests that
the later phases of the evolution of any given afterglow would
be more likely to exhibit signatures of a stellar-wind
environment. In all the $\sigma_w\ge 0.1$ bubbles, this
wind-like behavior becomes more pronounced the lower the value
of the cooling parameter $a_1$. Since (as noted above) the value of $n_{\rm
H,equiv}$ also exhibits a systematic dependence on this
parameter (it decreases with decreasing $a_1$), one may expect
afterglows with higher inferred ambient densities to
preferrentially indicate a uniform ISM-like environment if they
originate in such bubbles.

The inferred radii of afterglow shocks typically lie between
$\gtrsim 10^{17}\ {\rm cm}$ and $\lesssim 10^{18}\ {\rm cm}$ 
(e.g., \cite{P99}; \cite{CL00}). These values are consistent with the upper
limit on the bubble's outer radius (eq. [\ref{R_b}]) for 
supranova--GRB time delays of $\gtrsim 1\ {\rm yr}$ to 
$\lesssim 10\ {\rm yr}$. We can check on whether typical
afterglow-emitting shocks will 
still be relativistic by the time they reach the outer edge of
the bubble at $R_b$
by solving the adiabatic evolution equation
\begin{equation}\label{shock_evolve}
[\Gamma^2(r)-1]M_{\rm eq}(r)+[\Gamma(r)-1]M_0 = (\Gamma_0-1)M_0\
, \quad\quad\ R_s<r<R_b
\end{equation}
where
\begin{equation}\label{M_eq}
M_{\rm eq}(r)\equiv\int_{R_s}^r 4\pi R^2 n_{\rm H,equiv}(R) m_p dR
\end{equation}
and $M_0\equiv E/\Gamma_0c^2$ (e.g., van Paradijs et
al. 2000). The bottom panels in Figures \ref{fig4} and
\ref{fig5} show the results for the PWB
solutions presented in \S \ref{results} using a representative
value of $\Gamma_0$ and two plausible values of $E$. It is seen
that, in all cases, the GRB outflow decelerates rapidly after
entering the bubble, and in weakly cooling PWBs the Lorentz
factor of the afterglow-emitting gas is at most a few by the time the
shock reaches $R_b$ (and is effectively nonrelativistic for the
$E=10^{52}\ {\rm ergs}$ solutions). Only in the case of an energetic shock and
a strongly cooling bubble is $\Gamma(R_b)$ appreciable (but even
then it remains $\lesssim 10$). It is worth bearing in mind,
however, that, if the outflow is collimated with a small
opening half-angle $\theta_j$, then it will start to strongly
decelerate due to lateral spreading when its Lorentz factor
decreases to $\sim 1/\theta_j$ (e.g., \cite{SPH99}; \cite{R99}),
so that even the more energetic shocks could become
nonrelativistic while they are still inside the
PWB.\footnote{Mass loading of the bubble by evaporation of the
ejecta shell will further contribute to the lowering of $\Gamma$.}   
These results are again quite insensitive to the value of
$\sigma_w$ and to the choice of the imposed constraint; this is
not surprising in view of the fact that they depend mostly on
the behavior of $n_{\rm H,equiv}(\tilde{r})$, which exhibits a
similar trend. 

A GRB shock that reaches the supranova ejecta shell at $r=R_b$ with a
Lorentz factor $>1$ would be rapidly decelerated to subrelativistic
speeds since the rest-mass energy
of the shell ($\sim 2 \times 10^{55} M_{\rm ej,10}\ {\rm ergs}$, where
$M_{\rm ej,10}\equiv M_{\rm ej}/10 \, M_\odot$) is in
most cases much greater than the (equivalent isotropic) shock
energy $E$. The spectral characteristics of the forward shock
after it enters the shell could be evaluated once the dynamical
evolution of the shock is calculated.\footnote{Although the
behavior of the shock in both the highly relativistic and the Newtonian
limits had been considered in the literature, so far there has
been no published treatment of the transition between these two regimes.}
Besides the anticipated alterations in the spectral scaling laws, one
may expect the numerical values of the various physical parameters 
to undergo dramatic changes
as the shock moves from the interior of the bubble to the
ejecta shell: in particular, $n_{\rm H}$ would likely increase
by several orders of magnitude, whereas $\epsilon_e$ and
$\epsilon_B$ would probably decrease significantly. In
addition, the sudden deceleration would drive a relativistic
reverse shock into the GRB outflow, whose emission may have an important
effect. The overall outcome is likely to be a
discontinuous change in the shape and evolution of the observed spectrum.
Ramirez-Ruiz et al. (2001) modeled a somewhat similar situation
that may arise when a GRB shock that propagates in a stellar
wind encounters a density bump. They suggested that an encounter
of this type could induce a brightening and reddening of the
afterglow spectrum and might explain observations of such a
behavior in several sources. The situation
considered by Ramirez-Ruiz et al. (2001) differs, however,
from a PWB--SNR transition in that the
density contrast as well as the jumps in $\epsilon_e$ and
$\epsilon_B$ (which were assumed to be negligible in the ``bump
in a wind'' model) would typically be much larger in the latter case.
It would thus be interesting to carry out a detailed
investigation of the observational implications of the shock
encounter with a dense shell in the explicit context of the PWB model.

\subsection{Interpretation of X-Ray Features in the Supranova/PWB Model}
\label{features}

The SNR shell bounding the PWB could also manifest itself by imprinting
X-ray features on the GRB afterglow spectrum. Indeed, recent
detections of such features in several GRB sources have been argued
to provide strong support for the supranova scenario (e.g.,
\cite{L99}; \cite{P00}; \cite{V01}; \cite{A00}; \cite{L01};
\cite{BFD02}). To date, four GRB sources (GRB 970508, GRB 970828,
GRB 991216, GRB 000214), observed $\sim 8-40\ {\rm hr}$ after the burst, showed
emission features in their postburst X-ray spectrum, and one source
(GRB 990705) exhibited an absorption feature that disappeared $13\ {\rm s}$
after the onset of the burst. These features most likely represent Fe 
K$\alpha$ lines or an iron K edge, and their detection implies that a large
quantity ($\gtrsim 0.1\ M_\odot$) of pure iron is located in the vicinity
($r \lesssim 10^{16}\ {\rm cm}$) of the GRB source. Such a large iron mass
is most naturally produced in a supernova explosion, and the inferred distance
of the absorber indicates that the supernova event preceded the GRB by at least
several months, as expected in the supranova picture. The association with
a supranova is further strengthened by the argument (\cite{V01}) that the 
abundance
of $^{56}$Fe (the product of the radioactive decay of $^{56}$Ni and $^{56}$Co)
in supernova ejecta is not expected to become significant until $\sim 10^2$
days after the explosion, during which time the ejected gas in a source like
GRB 991216 (in which the observed line width is consistent with
an outflow speed $\sim 0.1\, c$; \cite{P00})
would have traveled to a distance $\gtrsim 10^{16}\ {\rm cm}$ from the origin.

We now proceed to discuss how
the observed X-ray features can be interpreted in the context of the supranova 
scenario, and we consider the implications of this interpretation to
the PWB afterglow model presented in this paper. We concentrate
on the specific example of GRB 991216, which allows us to
capitalize on the analysis already carried out on this object by Piro et al. 
(2000) and Vietri et al. (2001); our interpretation does, however, differ in
its details from the model favored by the latter authors. We approximate the
ejecta as a thin spherical shell of radius $R_{\rm ej}$ and
density $n_{\rm ej}$. Although the ejecta of a supernova that is
not associated with a
pulsar may be expected to fill the volume into which it expands, in the case
of an inflating PWB the ejecta will be swept up and compressed into a
dense shell (e.g., \cite{C77}). The acceleration of this shell
by the lower-density bubble gas would subject it to a
Rayleigh-Taylor instability, which could lead to clumping (see,
e.g., \cite{J98}). 
As we argue below, such clumping is consistent with the data for
GRB 991216.\footnote{The presence of a strongly clumped shell
was already inferred by Lazzati et al. (2000) in GRB 990705
from their analysis of the X-ray absorption feature in that
source.}

We assume that the emission is induced by continuum irradiation
from the central region that commences around the time of the burst
but is not necessarily confined to the solid angle of the GRB outflow.
The part of the shell that is observable to
us at time $t$ is limited by
light-travel effects, so that, for a source observed up to time 
$t_{\rm max}$, the solid angle $\Delta\Omega$ from which Fe emission is 
received is given by
\begin{equation}\label{delta_omega}
\frac{\Delta\Omega}{4\pi} = \frac{1-\cos{\theta_{\rm max}}}{2} = \frac{c 
t_{\rm max}}{2(1+z)R_{\rm ej}} = \frac{1.1\times 10^{15}\
{\rm cm}}{R_{\rm ej}}\ ,
\end{equation}
where the angle $\theta$ is measured with respect to the line of sight to the
origin, and where we substituted numerical values appropriate to GRB 991216
(redshift $z= 1.02$, $t_{\rm max}= 40.4\ {\rm hr}$).
Piro et al. (2000) identified the X-ray feature in GRB 991216 as an Fe XXVI
H$\alpha$ line (rest energy $6.97\ {\rm keV}$) with a FWHM (as quoted
in \cite{L01}) of $\sim 0.15\, c$. Since we attribute the emission to material
that moves toward the observer with a speed of that order, we favor an
identification with a lower-energy line, specifically  Fe XXV
He$\alpha$ (rest energy $6.7\ {\rm keV}$), but our results are not
sensitive to this choice.\footnote{An independent argument for
an identification of the X-ray feature in a source like GRB
991216 with the Fe XXV He$\alpha$ line was presented by
Ballantyne \& Ramirez-Ruiz (2001), who demonstrated that an Fe XXVI
H$\alpha$ line is unlikely to be observed because of the removal of photons
from the line core by Compton scattering.} Based on the photoionization models 
of Kallman \& McCray (1982), Fe XXV is the dominant ion when the
ionization parameter $\xi\equiv L_i/n_{\rm ej}R_{\rm ej}^2$ (where $L_i$ is the
ionizing continuum luminosity) lies in the range $\log{\xi} \sim 2.7-3.2$. 
Using $L_i=4\pi D^2 F_x\equiv L_{i,45}10^{45}\ {\rm ergs\ s^{-1}}$, with $F_x =
2.3\times 10^{-12}\ {\rm ergs\ cm^{-2}\ s^{-1}}$ and $D=4.7\ {\rm
Gpc}$ (\cite{P00}), we thus infer
\begin{equation}\label{xi}
n_{\rm ej}R_{\rm ej}^2 = 6.1\times 10^{42} (L_{i,45}/6.1)(\xi/10^3)^{-1}\, .
\end{equation}
The observed line luminosity corresponds to $\dot N_{\rm Fe,52} \equiv
(\dot N_{\rm Fe}/ 10^{52}\ {\rm photons\ s^{-1}})=8$ (\cite{P00}), and
we can write $\dot N_{\rm Fe}= (\Delta\Omega/4\pi)M_{\rm
Fe}/56m_pt_{\rm rec}$,  
where $M_{\rm Fe}=0.1M_{\rm Fe,0.1}\, M_\odot$ is the total
iron mass in the shell and
$t_{\rm rec} \approx 4\times 10^9T_6^{0.6}Z^{-2}n_e^{-1}=2.8\times
10^{10}T_6^{0.6}n_e^{-1}\ {\rm s}$ is the recombination time for a
$Z=24$ ion (with $T_e$ and $n_e$ being the electron temperature and
number density, respectively, and $Z$ the ion charge). The expression
for $t_{\rm rec}$ is valid in the range $T_e \sim 10^2-10^6\ {\rm K}$
(\cite{L01}), and photoionization models imply that $T_6\equiv
T_e/10^6\ {\rm K} \approx 1$ for $\log{\xi}\approx 3$ (\cite{KM82}).
Approximating $n_e\approx n_{\rm ej}$, we obtain
\begin{equation}\label{n_ej}
n_{\rm ej}= 1.0\times 10^9\, (4\pi/\Delta\Omega)T_6^{0.6}(\dot
N_{\rm Fe,52}/8)M_{\rm Fe,0.1}^{-1}\ {\rm cm^{-3}}\, .
\end{equation}

Substituting equation (\ref{n_ej}) into equation (\ref{xi}) gives
\begin{equation}\label{R_ej}
R_{\rm ej}=7.6\times 10^{16}(\Delta\Omega/4\pi)^{0.5}M_{\rm Fe,0.1}^{0.5}
(\dot N_{\rm Fe,52}/8)^{-0.5}(L_{i,45}/6.1)^{0.5}
T_6^{-0.3}(\xi/10^3)^{-0.5}\ {\rm cm}\, .
\end{equation}
Combining equations (\ref{R_ej}) and (\ref{delta_omega}) then yields
\begin{equation}\label{R_est}
R_{\rm ej}=1.9\times 10^{16}M_{\rm Fe,0.1}^{1/3}(\dot N_{\rm
Fe,52}/8)^{-1/3}(L_{i,45}/6.1)^{1/3} T_6^{-1/5}(\xi/10^3)^{-1/3}\ {\rm cm}
\end{equation}
and (for the given fiducial values) $\theta_{\rm max} \approx
28^\circ$. Substituting the estimate
(\ref{R_est}) into the relation (\ref{xi}) in turn implies
\begin{equation}\label{n_est}
n_{\rm ej} = 1.8\times 10^{10} M_{\rm Fe,0.1}^{-2/3}(\dot N_{\rm
Fe,52}/8)^{2/3}(L_{i,45}/6.1)^{1/3} T_6^{2/5}(\xi/10^3)^{-1/3}\ {\rm
cm^{-3}}\, .
\end{equation}
If the shell expands with a speed $v_{\rm ej}\approx 0.1c$ (see \S
\ref{scenario}), then its age when it reaches the radius given by
equation (\ref{R_est}) is
\begin{equation}\label{t_age}
t_{\rm age} \approx 72\, M_{\rm Fe,0.1}^{1/3}(\dot N_{\rm
Fe,52}/8)^{-1/3}(L_{i,45}/6.1)^{1/3}
T_6^{-1/5}(\xi/10^3)^{-1/3}(v_{\rm ej}/0.1 c)^{-1}\ {\rm days}\ .
\end{equation}
This value is consistent with the time required for
the bulk of the ejected radioactive $^{56}$Ni to decay into
$^{56}$Fe. 

The electron column density in the X-ray emitting portion of
the shell is given by $N_e = M_{\rm ej}/4\pi R_{\rm
ej}^2f_A\mu_em_p$, where $f_A$ is the covering factor of the
shell and $\mu_e$, the electron molecular weight, is 2 for
hydrogen-free ejecta. Assuming $f_A \approx 1$ and using
the estimate (\ref{R_est}), the Thomson optical depth of the shell is
inferred to be
\begin{equation}\label{tau_T}
\tau_{\rm T} = 0.9\, M_{\rm ej,10}M_{\rm Fe,0.1}^{-2/3}(\dot N_{\rm
Fe,52}/8)^{2/3}(L_{i,45}/6.1)^{-2/3}
T_6^{2/5}(\xi/10^3)^{2/3}(\mu_e/2)^{-1}\ .
\end{equation} 
For these fiducial values, the thickness of a homogeneous shell
would be $\sim 8\times 10^{13}\ {\rm cm}$, which is consistently $\ll R_{\rm ej}$. 
It is, however, more likely that this nominal thickness
corresponds to the size of a clump in a shell with a
small volume filling factor (see \cite{L01}). In fact, a high
degree of clumping is also indicated by the
requirement that the line photons reach the observer without
undrgoing excessive Compton broadening in the shell. The
photoionization optical depth of the iron ions in the shell is
similarly inferred to be
\begin{equation}\label{tau_Fe}
\tau_{\rm Fe} = 4.4\, M_{\rm Fe,0.1}^{1/3}(\dot N_{\rm
Fe,52}/8)^{2/3}(L_{i,45}/6.1)^{-2/3}
T_6^{2/5}(\xi/10^3)^{2/3}(\eta/0.5)\ ,
\end{equation} 
where $\eta$ is the relative abundance of the Fe XXV ion (e.g., \cite{KM82})
and where we used $\sigma_{\rm Fe XV}\approx 2.0\times 10^{-20}\ {\rm
cm^{-2}}$ (e.g., \cite{KK87}). (Our fiducial mass ratio $M_{\rm
Fe}/M_{\rm ej}=0.01$ corresponds to an iron abundance that is $\sim
5.6$ times the solar value.) The estimated values of $\tau_{\rm T}$
($\lesssim 1$) and $\tau_{\rm Fe}$ (a few) are optimal for producing
high--equivalent-width iron lines through reflection (e.g., \cite{WMKR00};
\cite{V01}). Since, in this picture, $\tau_{\rm Fe}\propto 1/R_{\rm ej}^2$,
the efficiency of producing detectable emission lines would typically
be low for shells with radii much in excess of $\sim 10^{16}\ {\rm
cm}$ (eq. [\ref{R_est}]). 

The most natural way of relating the above scenario to the PWB
model is to identify $t_{\rm age}$ with $t_{\rm sd}$ and $R_{\rm
ej}$ with $R_b$. However, such a straightforward identification
is problematic in that the magnitude of $R_{\rm ej}$ that is inferred
from the X-ray emission-line observations ($\lesssim 10^{16}\
{\rm cm}$; eq. [\ref{R_ej}]) is at least an order of magnitude
smaller than the lower limit on $R_b$ typically implied by the afterglow
data. In particular, in the case of GRB 991216, the optical
light curve showed evidence for steepening (which was attributed to shock
deceleration triggered by the lateral spreading of a jet)
starting about 2 days after the burst (\cite{H00}). For this
time scale to be consistent with an emission radius
$\sim 10^{16}\ {\rm cm}$, the relation $t \lesssim
(1+z)r/4c\Gamma^2$ implies that the flow Lorentz factor must be
$\lesssim 2$. However, given the comparatively high ($\sim
10^{53}-10^{54}\ {\rm ergs}$) equivalent isotropic energy
inferred for the emitting shock, it is unlikely that the Lorentz
factor could become so low over such a relatively short distance
(see bottom panels in Figs. \ref{fig4} and \ref{fig5}). The
problem is even more acute for
GRB 970508, in which the X-ray emission feature detected $\sim 1\ {\rm
day}$ after the burst again implies an emission radius $\sim
10^{16}\ {\rm cm}$ (e.g., Lazzati et al. 1999), but where
model fitting of the afterglow spectrum $\sim 1\ {\rm week}$ after the GRB
yields a radial scale $\gtrsim 3\times 10^{17}\ {\rm cm}$ along the line
of sight to the center [see references in \S \ref{introduction};
this result is supported by radio scintillation measurements
(\cite{F97})]. These values are mutually inconsistent, since the
SNR shell could not have reached a distance of $\gtrsim 0.1\ {\rm
pc}$ in one week even if it expanded at the speed of light. As
we noted in \S \ref{parameters}, the afterglow emitting gas
should decelerate rapidly after the forward shock encounters
the SNR shell, and the shock transition into the shell would
likely result in a discontinuous variation in the afterglow
light curve. If the radius of the
shell indeed coresponds to the value of $R_{\rm ej}$ indicated by
the X-ray emission-line data, then this is hard to reconcile with the
fact that, in the case of GRB 970508, the light curve
remained detectable and more or less smooth during a 450-day
monitoring period, with the underlying flow evidently becoming
nonrelativistic only after $\sim 100\ {\rm days}$ (\cite{FWK00};
see also \cite{CL00}).

The discrepancy between the inferred radius of the X-ray
line-emitting shell and the deduced radial distance of the
afterglow-emitting shock may be reconciled within the
framework of the supranova/PWB scenario if the SNR shell and the
PWB are not spherically symmetric. One possibility (suggested
by Lazzati et al. 1999 and \cite{V01}) is that the supernova
explosion does not eject matter along the SMNS rotation axis,
where the GRB outflow is subsequently concentrated. An
alternative possibility (which we discuss below) is that both
the SNR and the PWB become elongated in the polar directions
because of a preexisting density anisotropy in the GRB
environment. Under these circumstances, a highly collimated GRB outflow (such
as the one inferred in GRB 991216; \cite{H00}) could reach a
distance $\gtrsim 10^{17}\ {\rm cm}$ without encountering the
SNR shell even as the lower-latitude regions of the shell (from
which the X-ray line emission emanates) have radii $\lesssim
10^{16}\ {\rm cm}$. In the case of GRB 991216, where the X-ray
observations lasted between 37 and $40.4\ {\rm hr}$ after the
burst (\cite{P00}), the inferred effective spherical radius of the X-ray
emitting shell (eq. [\ref{R_ej}]) strictly corresponds only to
angles $\theta$ that lie in the narrow range $\sim 27-28^\circ$
(see eq. [\ref{delta_omega}]). If the jet opening half-angle is
significantly smaller than these values and $R_b$ is $\gg
10^{16}\ {\rm cm}$ at small values of $\theta$,
then the afterglow observations can in principle be consistent
with the X-ray emission-line measurements.\footnote{The detection of an X-ray
absorption feature would be compatible with this interpretation
if it could be demonstrated that the absorbing material was also
located at a distance $\gg 10^{16}\ {\rm cm}$ from the
irradiating-continuum source. In the only such case reported to
date (GRB 990705), Lazzati et al. (2001) deduced a radius $\sim
10^{16}\ {\rm cm}$ using a similar scheme to the one applied
here to the interpretation of X-ray line emission. They have,
however, also argued that the afterglow emission
properties in this object may be consistent with a shock/SNR-shell
encounter on this radial scale.}

The formation of a highly elongated PWB in the supranova
scenario may be a natural outcome of the manner in which its
environment was shaped by the progenitor star as well as of
its intrinsic physical properties. The star that gave rise to an
SMNS in a supranova event must have been massive, rapidly
rotating, and magnetized. It would have influenced the density
distribution in its vicinity through episodes of strong mass
loss, in particular during its red-supergiant and
blue-supergiant evolutionary phases. There is strong
observational evidence that the ``slow'' red-supergiant wind is often
anisotropic (possibly as a result of fast-rotation and magnetic
effects), transporting significantly more mass near the
equatorial plane than in the
polar regions. Subsequent stellar outflows that propagate into
this mass distribution will assume an elongated morphology: this
has been the basis of the ``interacting
stellar winds'' class of models for the shapes of planetary nebulae (e.g.,
\cite{DCB96}), in which the outflow represents the ``fast''
blue-supergiant wind, as well as of models of apparent SNR
``protrusions,''  in which the outflow corresponds to the
supernova ejecta (e.g., \cite{BLC96}). In these applications,
the subsequent outflows have been taken to be effectively
spherically symmetric. However, an even stronger collimation is
achieved if the later outflow is itself anisotropic. In
particular, if the fast wind is even weakly magnetized (with a dominant 
azimuthal field component), then, after passing through the wind
shock where the field is amplified (an effect that will be
especially strong if cooling is important behind the shock), the
magnetic hoop stress
will collimate the resulting interstellar bubble (e.g., \cite{CL94}). In
fact, as was argued by Gardiner \& Frank (2001), the collimation may
start even before the shock is encountered; this should be
particularly pronounced in cases where magnetic stresses also play a
dominant role in driving the fast wind (as in the Wolf-Rayet
wind model of \cite{BC93}). The additional collimation provided
by the magnetic field was suggested as the origin of strongly elongated
planetary nebulae, which cannot be readily explained by purely
hydrodynamic models.

A pulsar wind expanding into the anisotropic density
distribution created by the earlier (red-supergiant and
blue-supergiant) stellar outflows will give rise to an elongated
bubble (see, e.g., \cite{LB92} for a discussion of PWB evolution
in a stratified medium). Furthermore, since the pulsar wind
is highly magnetized and cooling may be important in the
supranova-induced PWB (see \S \ref{results}), the same magnetic collimation
effects that are invoked in the modeling of planetary nebulae
will act to increase the bubble elongation in this case
too. [In fact, the collimating effect of magnetic hoop stresses
on plerionic supernova remnants was already discussed by Rees \&
Gunn (1974); it was subsequently modeled by Begelman \& Li
(1992).] Under these combined effects, it is quite plausible to
expect that a bubble aspect ratio $\gtrsim 10$ could be
achieved, although this needs to be confirmed by an explicit
calculation.\footnote{In this case the width and centroid
redshift of the observed X-ray emission lines may not be due
entirely to the bulk motion of the SNR shell but may also have
significant contributions from Compton broadening within the
shell; see \cite{V01}.} Previous numerical simulations of
outflows expanding into an anisotropic medium also make it
likely that the column density
of the swept-up SNR shell will be lower near the apex of the
bubble than at larger values of $\theta$, which should be
relevant to the modeling of X-ray absorption and emission
features as well as of the afterglow light curve.
The expected departure of the PWB from sphericity might
require a modification of the model presented in \S \ref{structure},
which would probably be best done with guidance from numerical
simulations. We nevertheless anticipate that the results obtained from
the semianalytic model would remain at least qualitatively
valid. Furthermore, if a strong elongation only occurs near
the symmetry axis (which would be consistent with the data for
GRB 991216 as well as with some of the existing numerical
simulations), then even the quantitative predictions of the
simple spherical model would still be approximately correct.

\section{Conclusions}
\label{conclusions}
We propose to identify the environment into which
afterglow-emitting shocks in at least some GRB sources propagate
with pulsar-wind bubbles. Our results can be summarized
as follows:
\begin{itemize}
\item PWBs provide a natural resolution of
the apparent difficulty of accounting for the high
electron and magnetic energy fractions
($\epsilon_e$ and $\epsilon_B$, respectively) inferred in a
number of afterglow sources. This is because pulsar
winds are expected to have a significant $e^\pm$ component and
to be highly magnetized. If high values of $\epsilon_e$ in fact
prove to occur commonly in afterglow sources,
then this would strengthen the case for a simple,
``universal'' explanation of this type.
\item An association of PWBs with GRBs is expected under several
GRB formation scenarios, including
the collapse of a massive star.
In light of suggestive evidence that many of the afterglows observed to
date may have a massive stellar progenitor, we have concentrated
on this case. In particular, we considered the supranova
scenario of VS, in which an intense pulsar-type wind from the
GRB progenitor is a key ingredient of the hypothesized
evolution. In this picture, the ejection of a highly energetic, 
ultrarelativistic pulsar wind is predicted to follow the
supernova explosion and to last anywhere from several months to several years
until the central object collapses to form a black hole, thereby
triggering the burst. Recent detections of X-ray features in
several GRB sources have been interpreted as providing strong
support for this scenario.
\item To assess the implications of a PWB environment to
afterglow sources in the context of the supranova scenario, we
have constructed a simple, steady-state model of the early-time
structure of a plerionic supernova remnant. 
We have been guided by Atoyan's (1999) spectral modeling of the
Crab, which yielded a
lower initial wind Lorentz factor and a higher initial pulsar
rotation rate than in previous estimates, and by other recent
results on the Crab and Vela synchrotron nebulae, from which
we inferred a plausible range of the wind magnetization
parameter $\sigma_w$ ($\sim 10^{-3}-1$). In contradistinction to
previous models of the
structure of plerionic SNRs, we have replaced the assumption
that ideal MHD applies throughout the PWB with the postulate
that the electromagnetic-to-thermal pressure ratio in the bubble
remains constant after it increases to $\sim 1$. We have also
explicitly incorporated synchrotron-radiation cooling.
Although our solutions do not provide an exact representation of
radiative (and thus intrinsically time-dependent) PWBs, we have
verified that they generally do not depend on the detailed
approximations that are adopted and are essentially
characterized by $\sigma_w$ and by a
second parameter that measures the relative importance of
radiative cooling within the bubble.
It would be of interest to further develop this model and to
investigate the possibility that it can be applied both to young
radio pulsars and to GRB progenitors as members of the same general
class of rapidly rotating and strongly magnetized neutron stars.
\item In view of the ``hot'' (relativistic) equation of state
and high magnetization of the shocked wind, the effective
hydrogen number density that determines the properties of a
relativistic afterglow-emitting shock is given by $n_{\rm H, equiv}=
[4p+(B^\prime+{\mathcal{E}}^\prime)^2/4\pi]/ m_p c^2$, where
$B^\prime$ and ${\mathcal{E}}^\prime$ are, respectively, the
comoving magnetic and electric fields and $p$ is the particle
pressure. For plausible values of the cooling parameter
(and independent of the value of $\sigma_w$), the derived values
of $n_{\rm H, equiv}$ span the
density range inferred from spectral modeling of GRB
afterglows. An interesting
feature of the solutions is the predicted radial variation of 
$n_{\rm H, equiv}$ within the bubble, which can mimic either a
uniform-ISM or a stellar-wind environment, but which in general
exhibits a more diverse behavior.
Among other things, this model makes it possible to understand
how a GRB with a massive progenitor can produce an afterglow
that shows no evidence of a stellar-wind or a high-density
environment.
\item We have examined the dependence of the characteristic
synchrotron spectral quantities in an afterglow-emitting shock
that propagates inside a PWB on the bubble parameters and
related them to the standard expressions derived under the
assumption of a uniform-ISM environment. We found that, under
typical circumstances, the standard expressions remain roughly
applicable if one substitutes for $\epsilon_e$, $\epsilon_B$, and
$n_{\rm H}$ their ``equivalent'' PWB expressions. 
We noted, however, that the parameter scaling laws would change
in strongly radiative bubbles: these differences might be detectable
in objects with high inferred ambient densities.
\item Finally, we considered the possible observational
manifestations of the dense supranova shell that surrounds the
PWB in this picture. In particular, we discussed how the X-ray
emission features
detected in objects like GRB 991216 may be interpreted
in the context of a supranova-generated PWB. We concluded that
both the X-ray features and the afterglow emission could be
explained by this model if the PWB were elongated, and we argued
that such a shape might be brought about by
anisotropic mass outflows from the GRB progenitor star.
\end{itemize}

\acknowledgments We are grateful to N. Vlahakis for pointing out
to us the need to account for the comoving electric field in
the PWB structure equations, and to the anonymous referee for encouraging us to
broaden the parameter range covered by our solutions. We also
thank J. Arons, J. Blondin, A. Frank, A. Heger, D. Lamb, C. Litwin, A. Olinto,
T. Piran, E. Ramirez-Ruiz, M. Rees, M. Vietri, and S. Woosley
for useful conversations or correspondence. AK acknowledges a
Forchheimer Fellowship at the Hebrew University, where this work
was begun. JG acknowledges a Priscilla and Steven Kersten
Fellowship at the University of Chicago. This
research was supported in part by NASA grant NAG 5-9063 (AK) and
by NSF grant PHY-0070928 (JG).

\newpage

\end{document}